\documentclass[pmlr,twocolumn,10pt]{jmlr} 

\usepackage{booktabs}
\usepackage{siunitx}
\usepackage{multirow}
\usepackage[table]{xcolor}
\usepackage{tikz}

\usepackage[switch]{lineno}

\theorembodyfont{\upshape}
\theoremheaderfont{\scshape}
\theorempostheader{:}
\theoremsep{\newline}

\jmlrvolume{}
\jmlryear{2026}
\jmlrsubmitted{}
\jmlrpublished{}
\jmlrworkshop{Under Review} 

\title[VoxCare: Studying Natural Communication Behaviors of Hospital Caregivers through Wearable Sensing]{VoxCare: Studying Natural Communication Behaviors of Hospital Caregivers through Wearable Sensing of Egocentric Audio}

\author{%
\Name{Tiantian Feng} \Email{tiantiaf@usc.edu}\\
\Name{Kleanthis Avramidis} \Email{avramidi@usc.edu}\\
\Name{Anfeng Xu} \Email{anfengxu@usc.edu}\\
\Name{Deqi Wang} \Email{deqiwang@usc.edu}\\
\addr University of Southern California, USA
\AND
\Name{Brandon M Booth} \Email{Brandon.M.Booth@memphis.edu}\\
\addr The University of Memphis, USA
\AND
\Name{Shrikanth Narayanan} \Email{shri@usc.edu}\\
\addr University of Southern California, USA
}

\begin{document}

\maketitle

\begin{abstract}
Healthcare professionals work in complex, high-stakes environments where effective communication is critical for care delivery, team coordination, and individual well-being. However, communication activity in everyday clinical settings remains challenging to measure and largely unexplored in human behavioral research. We present \texttt{VoxCare}, a scalable egocentric wearable audio sensing and computing system that captures natural communication behaviors of hospital professionals in real-world settings without storing raw audio. \texttt{VoxCare} performs real-time, on-device acoustic feature extraction and applies a speech foundation model–guided teacher–student framework to identify foreground speech activity. From these features, \texttt{VoxCare} derives interpretable behavioral measures of communication frequency, duration, and vocal arousal. Our analyses reveal how, when, and how often clinicians communicate across different shifts and working units, and suggest that communication activity reflects underlying workload and stress. By enabling continuous assessment of communication patterns in everyday contexts, this study provides data-driven approaches to understand the behaviors of healthcare providers and ultimately improve healthcare delivery.
\end{abstract}

\paragraph*{Data and Code Availability}
The data produced and processed during the current study are accessible upon request through our dataset website and the code for this work can be found in our GitHub repository.  Both links will be provided after peer review.



\paragraph*{Institutional Review Board (IRB)}
All study procedures were conducted in accordance with USC’s Health Sciences Campus Institutional Review Board (IRB) approval (study ID HS-17-00876), and all procedures adhered to IRB guidelines and the ethical principles set in the Declaration of Helsinki.

\section{Introduction}
\label{sec:introduction}
Major global health crises such as the COVID-19 pandemic have underscored the importance of investigating ways to support the well-being of workers, especially in high-stakes domains such as in healthcare.
For example, nursing, which lies at the core of the modern healthcare system, is facing unprecedented challenges and significant increases in work demand.
Nurses are required to work long shifts, typically $12$ hours in duration~\citep{richardson_12hour, dabner_12hour}.
Increasingly, they must handle an intense workload while providing patient care that requires constant vigilance and meticulousness.
In addition to the physical demands, nursing is also emotionally demanding, requiring the delivery of humane, empathetic, and culturally sensitive care over extended periods~\citep{cricco2014need}.
Consequently, healthcare professionals are particularly susceptible to stress, anxiety, and related negative psychological effects that can lead to burnout and job dissatisfaction~\citep{iacovides2003relationship, glass1993depression}, highlighting the need to investigate occupational behavior that contributes to these negative outcomes.

Notably, interpersonal communication is fundamental to healthcare delivery across domains, including prevention, diagnosis, treatment, therapy, rehabilitation, and education~\citep{fakhr2011exploring}.
Moreover, substantial scientific evidence has shown that effective interpersonal communication is strongly associated with high-quality care~\citep{leonard2004human}.  
Interpersonal communication in healthcare typically includes spoken language, gestures, and digital modalities such as email and text messaging.
Among these, speech remains a primary way for healthcare professionals to communicate with others and express themselves during work.
Speech signal, in particular, contains objectively measurable acoustic properties that reflect not only linguistic content, but also prosody, tone, and indicators of mental and physical status.
As a result, understanding the speech activity in healthcare delivery can provide meaningful insights into the dynamic states, behaviors, and well-being of healthcare professionals, as well as the nature and quality of their interactions within complex clinical environments.

Two fundamental dimensions of speaking patterns are the frequency and duration of conversational interactions. Prior research has shown that frequent communication is critical for developing group cohesiveness, coordinating work activities, and increasing interpersonal trust within teams~\citep{walther2005rules}.
Increased interpersonal communication in healthcare settings has also been associated with improved task accuracy~\citep{gordon2011unit}.
Moreover, emotional cues such as arousal phenomena, also referred to as activation and excitation by~\citet{arousal_berlyne}, from speech signals are another valuable marker to assess speaking patterns~\citep{bone2014robust}. The arousal phenomenon is supported by the fact that affective arousal changes involve physiological reactions, which in turn modify the process by which speech is produced~\citep{scherer1986vocal}. In general, prosodic features of speech, such as pitch and intensity, can reflect the speaker's arousal~\citep{bone2012robust, scherer1986vocal}. Like other affective constructs, arousal is critical to motivation, personality, and work productivity~\citep{wilson1990personality, arousal_manage_barsade, johnson1979arousal}.

\begin{figure}[ht] {
    \centering
    
    \includegraphics[width=0.82\linewidth]{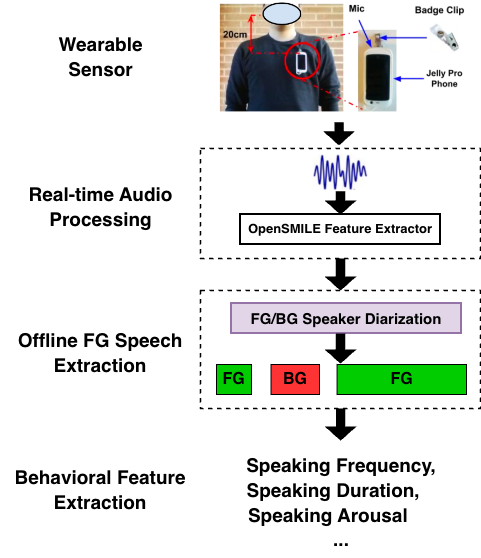}
    \vspace{-4mm}

    \caption{Overview of the \texttt{VoxCare}. \texttt{VoxCare} consists of a scalable wearable audio sensor that decodes speech signals to acoustic features in real time, an offline foreground/background speaker diarization, and a behavioral feature extractor.}
    \label{fig:egovox}
    \vspace{-3.5mm}
} \end{figure}

In this work, we introduce \texttt{VoxCare}, a scalable egocentric wearable audio sensing and computing system designed to capture natural interpersonal communication behaviors in occupational environments. Specifically, we deploy \texttt{VoxCare} in hospital settings to study everyday communication patterns among healthcare professionals, including nurses, laboratory staff, and physicians. The system integrates a lightweight wearable audio sensor that performs real-time, on-device processing to transform raw audio signals into low-level acoustic properties. To study communication from the wearer’s perspective, we develop a speaker detection model guided by a speech foundation model that identifies speaking regions corresponding to individuals wearing the sensor. Based on the retained acoustic features, \texttt{VoxCare} derives interpretable behavioral measures, such as speaking frequency and speech arousal, that characterize communication activity. 
Based on a 10-week longitudinal deployment, we provide quantitative analyses of everyday communication patterns among healthcare professionals and show their associations with self-reported work behavior, anxiety, and affect.

\vspace{-2mm}

\section{VoxCare}
\label{sec:methods}

Our proposed system, as shown in Figure~\ref{fig:egovox}, integrates a wearable audio sensing solution for real-time acoustic feature processing, an offline foreground speaker detection module to identify regions of interest, and a pipeline for extracting communicative behavioral features, including speaking frequency, speaking duration, and speaking arousal.

\vspace{-3mm}
\subsection{Wearable Audio Sensor}

\texttt{VoxCare} uses an unobtrusive wearable audio sensing system~\citep{feng2018tiles} that extracts descriptive features from raw speech in naturalistic settings. The system operates on a small, lightweight, and cost-effective Android device called the Jelly Pro smartphone~\citep{Jelly}, as presented in Fig~\ref{fig:egovox}. 
The audio sensing application running on this device integrates the openSMILE feature extraction toolkit~\citep{eyben2010opensmile}, which enables the decoding of a broad set of low-level acoustic features directly from raw audio. 
These features, including F0 (fundamental frequency), loudness, and a small set of filterbank features (MFCC 1-12), have been widely used to characterize speech patterns and speaking states. 
The sensing application runs at one-minute intervals and activates a 20-second feature extraction window to process and save acoustic features on the device. This choice is to achieve a balance between minimizing battery consumption and maximizing data collection.

Compared to previous wearable audio sensing solutions, such as EARS~\citep{mehl2001electronically} and SoundSense~\citep{lu2009soundsense}, our system offers improved privacy by transforming raw speech signals into compact acoustic features, which enables the interpretation of communication patterns without exposing sensitive language content. This is particularly critical for scaling deployment in naturalistic settings, where privacy concerns are universally present. The complete wearable system is highly cost‑effective, at roughly 100~USD per device. \texttt{VoxCare} is designed as a compact wearable for real‑world deployment, capable of scaling and adapting to researchers' needs by leveraging off‑the‑shelf Android hardware.

\subsection{Foreground Speaker Diarization}

\subsubsection{Problem Formulation} 

Given that the primary goal of \texttt{VoxCare} is to track and characterize communication activity in egocentric settings, it is critical to extract acoustic features that only originate from the speaker of interest. Similar to previous work presented by \cite{nadarajan2019speaker} and \cite{liang2023dataset}, we define \textit{foreground} speech activity as the vocal behavior produced by the speaker of interest (here, the sensor wearer), and \textit{background} speakers as all other speakers present in the environment. To facilitate the analysis of communication behaviors among healthcare professionals, we design an offline deep learning model to retain audio feature samples that belong to the foreground speech activity. Similar to speaker diarization that predicts the speaker in each audio frame, the deep learning model predicts each frame of audio features as foreground speech (FG), background speech (BG), or silence and environmental noise (S).

\begin{figure*} {
    \centering
    
    \includegraphics[width=0.88\linewidth]{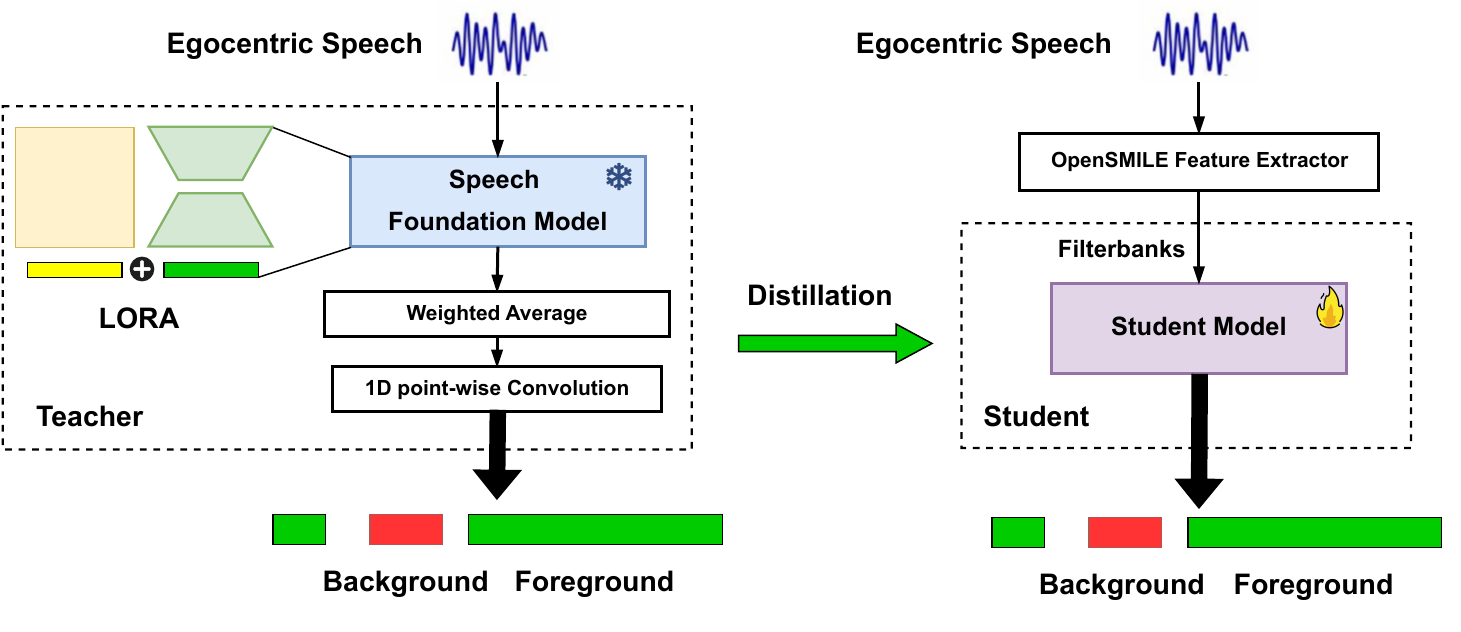}
    \vspace{-5mm}
    \caption{Overview of the foreground speaker diarization process. 
    \texttt{VoxCare} uses a teacher–student learning framework for foreground speaker diarization. Specifically, a teacher network based on a speech foundation model is first trained to perform foreground speaker diarization. The teacher network is then transferred to a lightweight student network through knowledge distillation.}
    \label{fig:egovox_fg}
    \vspace{-3.5mm}
} \end{figure*}

\vspace{-3mm}
\subsubsection{Training Data Selection}
Acquiring raw audio waveforms from hospital environments to curate an annotated training corpus was infeasible due to privacy and institutional constraints. Hence, we trained the proposed FG/BG speaker diarization model using publicly available corpora that contain egocentric speech collected in a similar form factor. In particular, we used four publicly available datasets in our experiments, namely MMCSG~\citep{zmolikova2024chime}, the ICSI-Meeting Corpus~\citep{janin2003icsi}, the EasyCom Corpus~\citep{donley2021easycom}, and an Internal Meeting Corpus collected with the same Jelly Pro smartphone. 

\vspace{-3mm}
\subsubsection{Teacher-student Distillation} One key constraint of the proposed audio sensing system is the need to operate with a highly compact feature representation, which is driven by considerations of data sensitivity, on-device computation, and limited storage capacity. As a result, only a small set of acoustic features, such as filterbanks (MFCC 1-12), is available for modeling and inferring foreground speech regions. Within this constrained feature space, we propose a lightweight foreground/background (FG/BG) speaker diarization module that is trained via knowledge distillation from a teacher foreground speaker diarization model built upon a speech foundation model.

\vspace{1mm}
\noindent \textbf{Teacher Model} We first built a FG/BG speaker diarization benchmark using speech foundation models to create the teacher model. The considered foundation models include Whisper family models. Particularly, all models were trained using low-rank adaptation (LoRA, by~\citet{hu2022lora}) and used a unified architecture consisting of a weighted average layer that combined all hidden layers and a 1D convolutional neural network that mapped the weighted output to frame-wise FG/BG speaker prediction.

\vspace{1mm}
\noindent \textbf{Student Model} In parallel, we created a lightweight baseline model that operates directly on filterbank features to perform FG/BG speaker diarization as the student model.
Specifically, we used the ResNet models in our experiments to create the student model and predict frame-wise FG/BG speaker labels.

\vspace{1mm}
\noindent \textbf{Teach-student Distillation}
Based on empirical performance, we then selected the best-performing teacher–student pair to initialize the second stage of knowledge distillation training. 
In the distillation process, the weights of the teacher model were frozen while we continued fine-tuning the student model. The training objective combined a frame-wise cross-entropy loss for classification with a KL-divergence loss that encourages the student's outputs to align with those of the teacher, as shown below.

\begin{equation}
    \mathcal{L}_{total} = \mathcal{L}_{CE}(p_{s}) + \alpha \mathcal{L}_{KLD}(p_{t}, p_{s})
    \label{equ:teacher_student_loss}
\end{equation}

This approach enables the development of an effective FG/BG speaker diarization model that operates entirely on low-dimensional acoustic features.

\begin{figure} {
    \centering
    
    \includegraphics[width=0.95\linewidth]{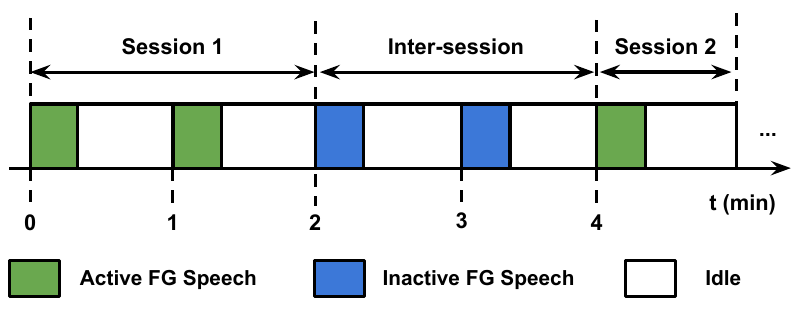}
    \vspace{-5mm}
    
    \caption{Example of the FG speech sessions. The x-axis represents the time in minutes. The 1st and the 2nd sessions contain the FG speech at $t={0, 1}$ and $t={4}$, respectively.}
    \vspace{-3.5mm}
    \label{fig:session_def}
} \end{figure}

\subsection{Communication Behavior Extraction}

Here, we describe how communication features are computed. Since the wearable audio sensors perform real-time speech processing for 20 seconds every minute, we therefore define a \textit{speech session} as a sequence of discrete recordings with an offset of one minute between each consecutive recording (the recording also must have at least $200$ foreground speech frames). Figure \ref{fig:session_def} shows an example speech session, where a set of recordings are at time $\mathbf{t}=\{0, 1, 4\}$, and we can group these recordings into two sessions at a time $\mathbf{t_{1}}=\{0, 1\}$ and at time $\mathbf{t_{2}}=\{4\}$.

\subsubsection{Speaking Frequency and Duration}

Based on the definition of speech sessions, we computed the following features to quantify the speaking frequency and duration. First, we computed the speaking sessions per hour. We do not consider total speaking sessions, given that different positions work varying duration which can introduce systematic differences. This feature describes how frequently a participant interacts with others. Moreover, we computed the average session duration to characterize the average speaking duration. For each participant, features were averaged across all recorded shifts.

\vspace{-3mm}
\subsubsection{Speaking Arousal}

In addition to speaking frequency and duration, we applied a rule-based arousal estimation framework~\citep{bone2014robust} to estimate how healthcare professionals communicate during work. This method has been validated to achieve a strong correlation between estimated vocal arousal scores and ground-truth arousal ratings in different contexts. Particularly, this approach is generalizable and favorable when collecting raw training samples is not feasible. This method only requires small sets of acoustic features, without the need for emotion labels. 

\vspace{1mm}
\noindent \textbf{Arousal Estimation from a Single Feature} The feature sets used in this study were log-pitch, intensity, and the ratio of high-frequency to low-frequency energy. First, this framework builds an empirical distribution $N_{i}$ for each feature type $i\in\{1, 2, 3\}$. As~\citet{bone2014robust} suggested, we create personalized empirical models to account for differences in speaker characteristics. Then, the arousal intensity score $p_{i, j}$ is computed as the median value of feature type $i$ in the $j$-th 20-second recording with the corresponding empirical model, $N_{i}$, by:
\begin{equation}  
    p_{i, j} = 2 \times E[x_{i, j} > N_{i}] - 1
    \label{equ:arousal}
\end{equation}

\noindent \textbf{Arousal Score Fusion} To gain a more robust arousal estimation, a fusion technique is applied to combine the scores from individual arousal ratings. The weights for fusion are computed as the Spearman's rank-correlation coefficient $r_{i}$ between each score vector $p_{i}$ and the score mean vector $p_{\mu}$, where the vectors are composed of scores for all of a speaker's utterances. We define the correlation vector as $\mathbf{r} = (r_{1}, r_{2}, r_{3})$, and the weights are then normalized to have unit combined magnitude: $w_{i} = \frac{r_{i}}{||\mathbf{r}||}$.

\vspace{0.5mm}
\noindent \textbf{Arousal Score Intepretation}
\citet{bone2014robust} suggest that, if annotated neutral affect baselines are established, the obtained arousal scores may be interpreted such that positive scores indicate more aroused (or active) states, with the magnitude being associated with confidence. In the absence of an annotated neutral baseline, we instead interpret positive arousal scores as more active states relative to each participant’s overall baseline. 
We derive the 90th percentile of the arousal score to describe the speaking arousal patterns in the 1st-half and 2nd-half of the work shift. Similar to speaking frequency and duration features, for each participant, we computed the average of each feature by averaging across all recorded shifts.

\vspace{-2mm}
\section{Study Data Collection}

\vspace{-2mm}
\subsection{Study Introduction}

Our wearable audio sensor was deployed as part of the TILES study~\cite{ mundnich2020tiles, yau2022tiles}, a longitudinal experiment designed to examine how physiological and behavioral factors influence job performance and employee well-being. 
Over a period of ten weeks, the study collected multimodal data from hospital professionals employed at a tertiary/quaternary academic medical center in Los Angeles, CA, who had volunteered to participate in this study. 
In addition to natural communication sampling, data collection included monitoring of continuous physiological signals (such as with Fitbit Charge 2 wristbands) and self-reported assessments. 
All participants provided informed consent before the study. The study protocol was approved by the Institutional Review Board, and all procedures adhered to the ethical principles. 
To the best of our knowledge, this deployment represents one of the first large-scale evaluations of a wearable audio sensing system conducted in a real-world clinical environment. The complete study involved over 300 hospital professionals across different roles, including nurses, laboratory staff, and clinical administrative personnel, while this analysis focuses on a subset of 255 participants who complied with wearing audio sensors.

\vspace{-2mm}
\subsection{Self-report Outcomes}

To assess both work-related behaviors and mental health of hospital professionals, we selected a set of self-reported surveys that were collected at the beginning of the study, including the State-Trait Anxiety Inventory (STAI)~\citep{spielberger1971state} and the In-role behavior (IRB)~\citep{williams1991job}. We aimed to explore the associations between communication patterns at work and these self-reported surveys. Details regarding these two surveys were described in Appendix~\ref{apd:survey}

\vspace{-2mm}
\subsection{Participants and Data Statistics}

\begin{figure}[ht] {
    \centering
    \includegraphics[width=0.9\linewidth]{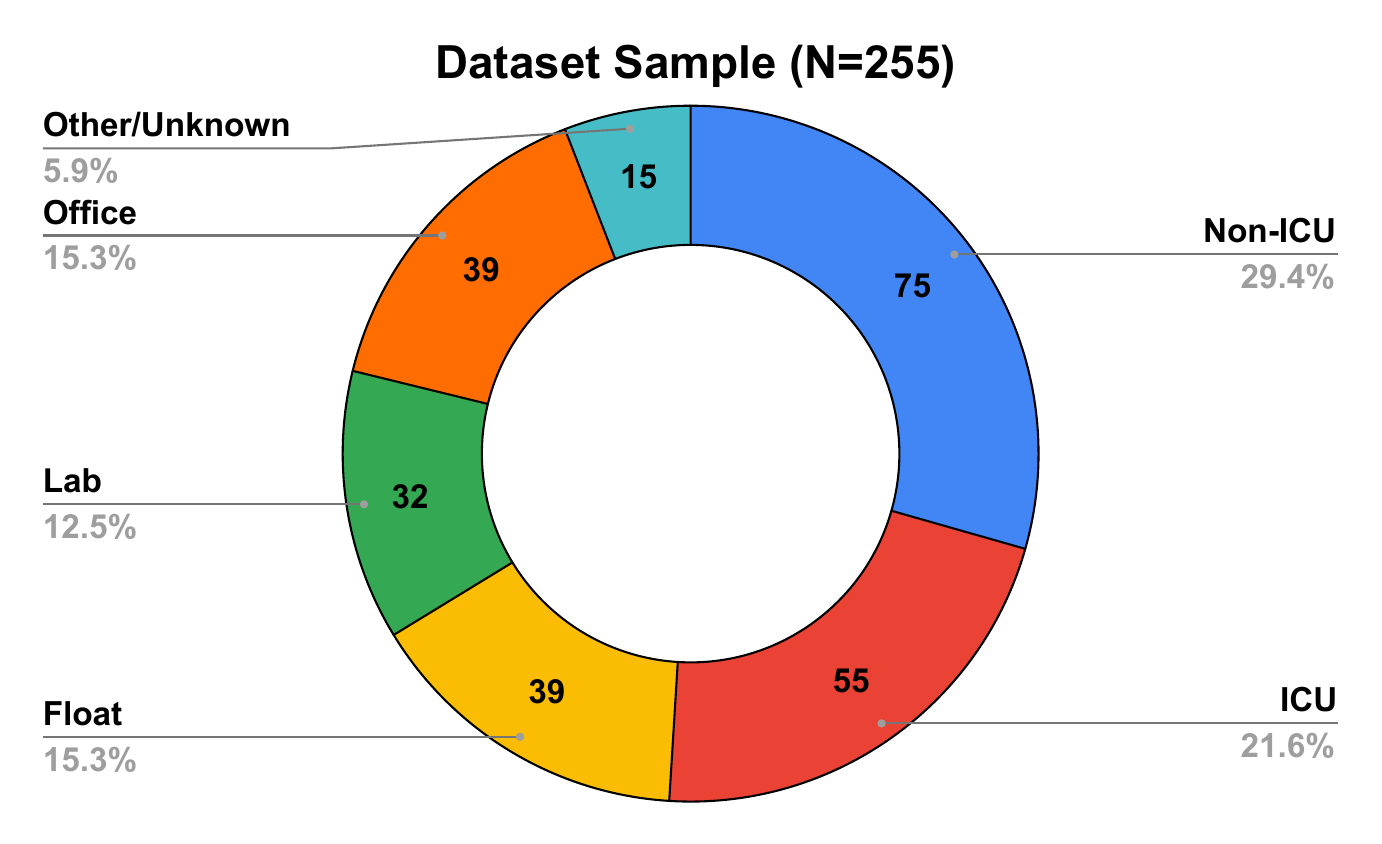} 
    \vspace{-5mm}
    \caption{Demographics of the primary working unit in the current dataset.}
    \label{fig:demo_job}
    \vspace{-3.5mm}
} \end{figure}

In total, 356 participants were enrolled in the study, of whom n=255 met the compliance criteria for inclusion in the current analysis by wearing the audio sensor. Minimum compliance was defined as having at least five recorded work shifts by the audio sensor per participant. Figure~\ref{fig:demo_job} visualizes the primary working unit within the current wearable audio dataset. Of the study participants, n=75 ($29.4\%$) worked in the non-ICU, n=55 ($21.6\%$) worked in the ICU, and n=39 ($15.3\%$) worked in the float pool. Notably, n=39 subjects ($15.3\%$) in the current cohort worked in office settings, including as administrators, doctors, or in research roles. Moreover, among 255 compliant participants, 182 ($71.4\%$) primarily worked day shifts, while the remaining 73 ($28.6\%$) worked night shifts.
The majority of participants are female ($n=180$, $70.6\%$) and the ages range from $25$ -- $65$, with n=130 ($51.0\%$) subjects $20$-$39$ years old, n=76 ($29.8\%$) subjects $40$-$49$ years old, and the remaining n=49 ($19.2\%$) subjects being over $50$ years old. 

\vspace{-1mm}
\section{FG Diarization Experiments}

\vspace{-0.5mm}
\subsection{Datasets} 
In these experiments, we used four egocentric audio datasets that were collected with a similar form factor. Specifically, both MMCSG~\citep{zmolikova2024chime} and the EasyCom corpus~\citep{donley2021easycom} were collected using smart glasses. The MMCSG corpus involves dyadic interactions recorded through smart Aria glasses worn by one of the subjects. The complete dataset includes 530 unique 3-minute recordings from 138 participants. Similarly, EasyCom includes recordings of 3--5 participants seated around a table for conversations of approximately 30 minutes. In total, there are 12 sessions with about 5 hours of audio recordings in the EasyCom dataset.

\begin{table}
    \footnotesize
    \centering

    \vspace{1mm}

    \scalebox{0.9}{
    \begin{tabular}{lccc}

        \toprule
        \multirow{1}{*}{\textbf{}} & 
        \multicolumn{1}{c}{\textbf{Nurse}} &
        \multicolumn{1}{c}{\textbf{Non-Nurse}} &
        \multicolumn{1}{c}{\textbf{Total}} \\

        \midrule

        \multicolumn{1}{l}{\textbf{Sex}} & & \\
        
        \hspace{0.2cm}{Male} &
        49 $(19.2\%)$ &
        26 $(10.2\%)$ &
        $75 (29.4\%)$ \\

        \hspace{0.2cm}{Female} &
        120 $(47.1\%)$ &
        60 $(23.5\%)$ &
        180 $(70.6\%)$ \\
        \midrule
        \multicolumn{1}{l}{\textbf{Primary Shift}} & & \\

        \hspace{0.2cm}{Day Shift} &
        105 $(41.2\%)$ &
        77 $(31.2\%)$ &
        $182$ $(72.2\%)$ \\

        \hspace{0.2cm}{Night Shift} &
        64 $(25.1\%)$ &
        9 $(3.7\%)$ &
        $73$ $(28.8\%)$ \\
        \midrule
        
        \multicolumn{1}{l}{\textbf{Age}} & & \\

        \hspace{0.2cm}{$<40$ Yr} &
        98 ($38.4\%$) &
        32 ($12.6\%$) &
        130 ($51.0\%$) \\

        \hspace{0.2cm}{$40-49$ Yr} &
        51 ($20.0\%$) &
        25 ($9.8\%$) & 
        76 ($29.8\%$) \\

        \hspace{0.2cm}{$\geq50$ Yr} &
        20 ($7.8\%$) &
        29 ($11.4\%$) &
        49 ($19.2\%$) \\

        \bottomrule

    \end{tabular}
    }
    
    \caption{Demographics of the primary work shift, sex and age in the current analysis.}
    \label{tab:demo_shift}
    \vspace{-4mm}

\end{table}

\begin{table*}[t]
    \centering

    \caption{Comparison of foreground speaker diarization results in \texttt{VoxCare}.}
    \vspace{2mm}
    \resizebox{\linewidth}{!}{
    \begin{tabular}{lcccccccccc}

        \toprule

        \multirow{3}{*}{\textbf{Model}} & 
        \multicolumn{4}{c}{\textbf{MMCSG}} & \multicolumn{4}{c}{\textbf{ICSI-Meeting Corpus}} 
        \\
         & \multicolumn{1}{c}{\textbf{Miss}} & \multicolumn{1}{c}{\textbf{False}} & \multicolumn{1}{c}{\textbf{Speaker}} & \multirow{2}{*}{\textbf{DER}} & \multicolumn{1}{c}{\textbf{Miss}} & \multicolumn{1}{c}{\textbf{False}} & \multicolumn{1}{c}{\textbf{Speaker}} & \multirow{2}{*}{\textbf{DER}}
        \\
         & \multicolumn{1}{c}{\textbf{Detection}} & \multicolumn{1}{c}{\textbf{Alarm}} & \multicolumn{1}{c}{\textbf{Confusion}} &  & \multicolumn{1}{c}{\textbf{Detection}} & \multicolumn{1}{c}{\textbf{Alarm}} & \multicolumn{1}{c}{\textbf{Confusion}} & 
        \\
        \cmidrule(lr){1-1}
        \cmidrule(lr){2-5}
        \cmidrule(lr){6-9}

        \rowcolor{gray!20}
        \textbf{Teacher Model} \\
        
        \hspace{4mm} Whisper-Tiny &
        9.8{\footnotesize$\pm$0.3} &  
        3.6{\footnotesize$\pm$0.2} &
        3.3{\footnotesize$\pm$0.3} & 16.8{\footnotesize$\pm$0.3} &
        10.5{\footnotesize$\pm$0.4} & 
        5.7{\footnotesize$\pm$0.3} & 
        1.4{\footnotesize$\pm$0.1} & 17.6{\footnotesize$\pm$0.1}
        
        \\

        \hspace{4mm} Whisper-Small &
        3.6{\footnotesize$\pm$0.1} & \textbf{3.4}{\footnotesize$\pm$0.1} & 0.8{\footnotesize$\pm$0.1} & \textbf{7.7}{\footnotesize$\pm$0.2} &
        6.0{\footnotesize$\pm$0.6} & 
        \textbf{3.9}{\footnotesize$\pm$0.5} & 0.2{\footnotesize$\pm$0.0} & 10.1{\footnotesize$\pm$0.1} \\

        \hspace{4mm} Whisper-Large &
        \textbf{3.5}{\footnotesize$\pm$0.2} & 3.5{\footnotesize$\pm$0.6} & \textbf{0.8}{\footnotesize$\pm$0.1} & 7.8{\footnotesize$\pm$0.5} &
        \textbf{4.3}{\footnotesize$\pm$1.1} & 
        4.5{\footnotesize$\pm$0.5} & 
        \textbf{0.2}{\footnotesize$\pm$0.1} & \textbf{9.0}{\footnotesize$\pm$0.6} \\

        \rowcolor{gray!20}
        \textbf{Student Model} \\

        \hspace{4mm} ResNet &
        6.3{\footnotesize$\pm$0.2} & 
        5.1{\footnotesize$\pm$0.8} & 
        \textbf{4.4}{\footnotesize$\pm$0.2} & 
        15.8{\footnotesize$\pm$1.0} &
        13.1{\footnotesize$\pm$0.6} & 
        9.3{\footnotesize$\pm$1.2} & 1.1{\footnotesize$\pm$0.1} & 23.4{\footnotesize$\pm$0.8} \\

        \hspace{4mm} \texttt{VoxCare}-ResNet (Ours) &
        \textbf{5.6}{\footnotesize$\pm$0.1} & 
        4.7{\footnotesize$\pm$0.1} & 
        \textbf{3.2}{\footnotesize$\pm$0.5} &
        \textbf{13.5}{\footnotesize$\pm$0.6} &
        \textbf{12.9}{\footnotesize$\pm$1.7} & 
        \textbf{7.9}{\footnotesize$\pm$1.5} & 
        \textbf{0.9}{\footnotesize$\pm$0.2} & 
        \textbf{21.7}{\footnotesize$\pm$0.6} \\

        \midrule
        \multirow{3}{*}{\textbf{Model}} & 
        \multicolumn{4}{c}{\textbf{EasyCom}} & \multicolumn{4}{c}{\textbf{Internal Meeting Corpus}} 
        \\
         & \multicolumn{1}{c}{\textbf{Miss}} & \multicolumn{1}{c}{\textbf{False}} & \multicolumn{1}{c}{\textbf{Speaker}} & \multirow{2}{*}{\textbf{DER}} & \multicolumn{1}{c}{\textbf{Miss}} & \multicolumn{1}{c}{\textbf{False}} & \multicolumn{1}{c}{\textbf{Speaker}} & \multirow{2}{*}{\textbf{DER}}
        \\
         & \multicolumn{1}{c}{\textbf{Detection}} & \multicolumn{1}{c}{\textbf{Alarm}} & \multicolumn{1}{c}{\textbf{Confusion}} &  & \multicolumn{1}{c}{\textbf{Detection}} & \multicolumn{1}{c}{\textbf{Alarm}} & \multicolumn{1}{c}{\textbf{Confusion}} & 
        \\
        \cmidrule(lr){1-1}
        \cmidrule(lr){2-5}
        \cmidrule(lr){6-9}

        \rowcolor{gray!20}
        \textbf{Teacher Model} \\
        \hspace{4mm} Whisper-Tiny &

        19.1{\footnotesize$\pm$0.3} & 9.4{\footnotesize$\pm$0.4} & 2.1{\footnotesize$\pm$0.1} & 30.6{\footnotesize$\pm$0.3} &

        13.4{\footnotesize$\pm$0.4} & 
        5.8{\footnotesize$\pm$0.6} & 
        6.8{\footnotesize$\pm$1.2} & 26.0{\footnotesize$\pm$1.3} \\

        \hspace{4mm} Whisper-Small &
        \textbf{17.7}{\footnotesize$\pm$0.3} & 
        4.6{\footnotesize$\pm$1.1} & 
        \textbf{1.0}{\footnotesize$\pm$0.1} & \textbf{23.3}{\footnotesize$\pm$0.8} &

        9.8{\footnotesize$\pm$1.4} & 
        \textbf{4.8}{\footnotesize$\pm$0.6} & 
        \textbf{1.1} \footnotesize$\pm$0.2 & 
        \textbf{15.6} \footnotesize$\pm$0.9 \\

        \hspace{4mm} Whisper-Large &
        18.2{\footnotesize$\pm$0.2} & \textbf{4.5}{\footnotesize$\pm$1.1} & 1.3{\footnotesize$\pm$1.0} & 24.0{\footnotesize$\pm$2.2} &

        \textbf{7.3}{\footnotesize$\pm$2.1} & 
        6.5{\footnotesize$\pm$1.4} & 
        2.0{\footnotesize$\pm$0.6} & 15.8{\footnotesize$\pm$0.3} \\

        \rowcolor{gray!20}
        \textbf{Student Model} \\

        \hspace{4mm} ResNet &
        24.5{\footnotesize$\pm$2.2} & 
        \textbf{3.5}{\footnotesize$\pm$1.2} & 
        2.5{\footnotesize$\pm$0.2} & 
        30.5{\footnotesize$\pm$1.1} &

        15.9{\footnotesize$\pm$1.9} & 
        7.2{\footnotesize$\pm$3.9} & 
        \textbf{7.6}{\footnotesize$\pm$0.7} & 30.7{\footnotesize$\pm$1.8} \\

        \hspace{4mm} \texttt{VoxCare}-ResNet (Ours) &
        \textbf{22.8}{\footnotesize$\pm$0.3} & 
        3.6{\footnotesize$\pm$0.3} & 
        \textbf{1.9}{\footnotesize$\pm$2.1} & 
        \textbf{28.4}{\footnotesize$\pm$0.1} &

        \textbf{14.6}{\footnotesize$\pm$2.3} & 
        \textbf{6.5}{\footnotesize$\pm$2.7} & 
        7.7{\footnotesize$\pm$0.8} & 
        \textbf{28.8}{\footnotesize$\pm$0.6} \\
        
        \bottomrule

    \end{tabular}
    }
    \label{tab:egovox}
    \vspace{-3mm}
\end{table*}

On the other hand, the ICSI meeting corpus~\citep{janin2003icsi} consists of natural multi-party conversations recorded with multiple microphones placed in the environment, enabling both close-talk and far-field recording. Moreover, we collected an internal meeting corpus in which colleagues wore the proposed wearable audio sensors during internal meetings. In total, there are 6 meetings (around 5 hours total) recorded from 12 participants. Manual annotations of speaker boundaries were available for each dataset to derive precise FG, BG, and silence regions.

\vspace{-3mm}
\subsection{Experimental Details}

All training experiments were conducted on a high-performance computing (HPC) cluster. We trained each model using a \textit{collective training set} that combines four datasets. Specifically, we used the default train/validation/test split for MMCSG. For the remaining datasets, we split recording sessions into 60\%, 20\%, and 20\% for training, validation, and testing, respectively. For each run, we used a learning rate of $5\times10^{-5}$ and trained for 15 epochs. $\alpha$ in Equation~\ref{equ:teacher_student_loss} was empirically tested at \{1, 5, 10\}, while we reported the best overall performance with $\alpha=5$.

Our teacher-model experiments included Whisper-Tiny, Whisper-Small, and Whisper-Large. Our student model was a ResNet~\citep{he2016deep} that takes MFCC 1--12 as input. Both models predict frame-wise FG/BG speaker labels in a 10-second window.
We first trained the teacher and student models independently, and then initialized the distillation stage from these pre-trained weights and continued training for 15 epochs. We selected the best teacher model in the distillation training. During distillation, the teacher model weights were frozen. We repeated each setup three times with different random seeds and report the average performance across runs. We report missed detection, false alarm, speaker confusion, and diarization error rate (DER), where lower values indicate better performance. Specifically, Missed Detection refers to the duration of actual speech (FG or BG) in a recording that the system fails to detect, while a false alarm occurs when the system incorrectly identifies non-speech audio, such as background noise or silence, as speech (FG or BG). 
We describe the error types (miss detection, false alarm, and speaker confusion) in Appendix~\ref{apd:fg_diarization}, while DER reflects the combined error rates.

\vspace{-1mm}
\subsection{Teacher Model Performance}

Table~\ref{tab:egovox} summarizes the teacher-model, student-model, and disllation model performance for FG/BG diarization. Overall, the Whisper family model yields strong performance in FG/BG diarization.
Across MMCSG, ICSI, EasyCom, and an Internal Meeting corpus, Whisper-Small and Whisper-Large consistently outperform Whisper-Tiny, yielding substantially lower miss and false-alarm rates, leading to overall lower diarization error rate (DER). In particular, Whisper-Small achieves the best DER on MMCSG ($7.7\%$), EasyCom ($23.3\%$), and the internal meeting dataset (15.6), while Whisper-Large provides the lowest DER on ICSI ($9.0\%$). Based on these results, we selected the Whisper-Small model as the teacher model for distillation training.

\vspace{-0.5mm}
\subsection{Student Model Performance}

In contrast, models that run directly on low-dimensional acoustic features (MFCC 1--12) show higher error rates, but they remain viable for on-device deployment after distillation. Specifically, the ResNet student baseline shows higher DER across all four corpora, whereas our distilled \texttt{VoxCare}-ResNet improves over ResNet consistently (e.g., MMCSG: $15.8\%\rightarrow13.5\%$, ICSI: $23.4\%\rightarrow21.7\%$, EasyCom: $30.5\%\rightarrow28.4\%$, Internal Meeting Data: $30.7\%\rightarrow28.8\%$). In particular, we observe that these gains are mainly driven by decreased speaker-confusion errors. Overall, these results suggest that distillation from a foundation-model teacher can improve FG/BG diarization when only a lightweight model can be deployed. Finally, we apply \texttt{VoxCare}-ResNet to infer FG/BG regions on real acoustic feature recordings collected from hospital professionals.

\vspace{-3mm}
\section{Communication Patterns of Hospital Professionals}

In this section, we analyzed communication patterns of hospital professionals across different working units and shifts within the [anonymized] study, using foreground speech regions inferred by the \texttt{VoxCare}-ResNet FG/BG diarization model shown in Table~\ref{tab:egovox}.

\subsection{Speaking Frequency and Duration}

We first compared the speaking session frequency and duration among different work shifts and work units. We applied three-way ANOVA in all statistical tests by adding sex and age as confounding factors.

\begin{figure}[ht] {
    \centering
    
    \begin{tikzpicture}
        
        \node[draw=none,fill=none] at (0, 3){\includegraphics[width=0.5\linewidth]{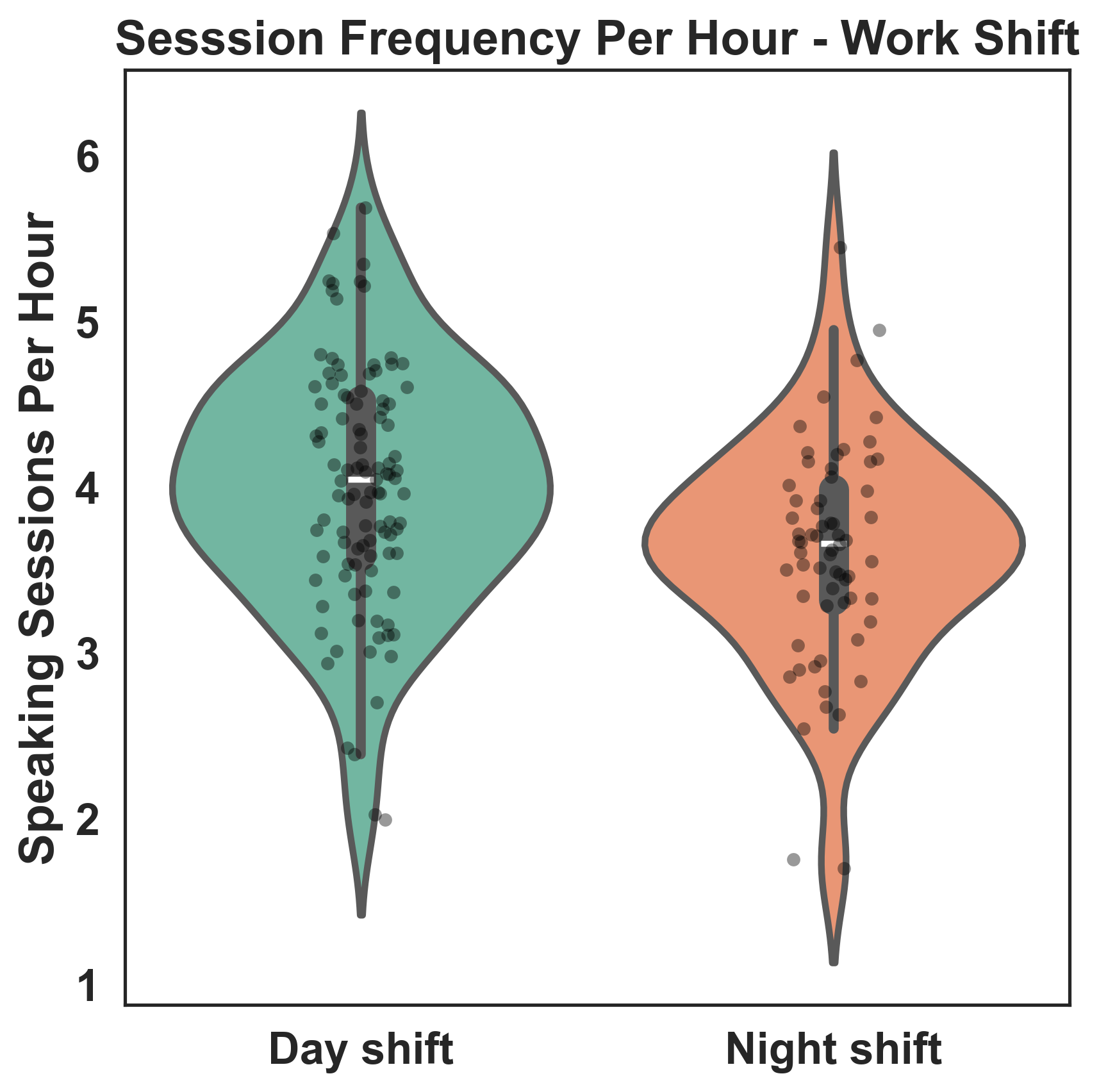}};
        
        \node[draw=none,fill=none] at (0.5\linewidth, 3){\includegraphics[width=0.5\linewidth]{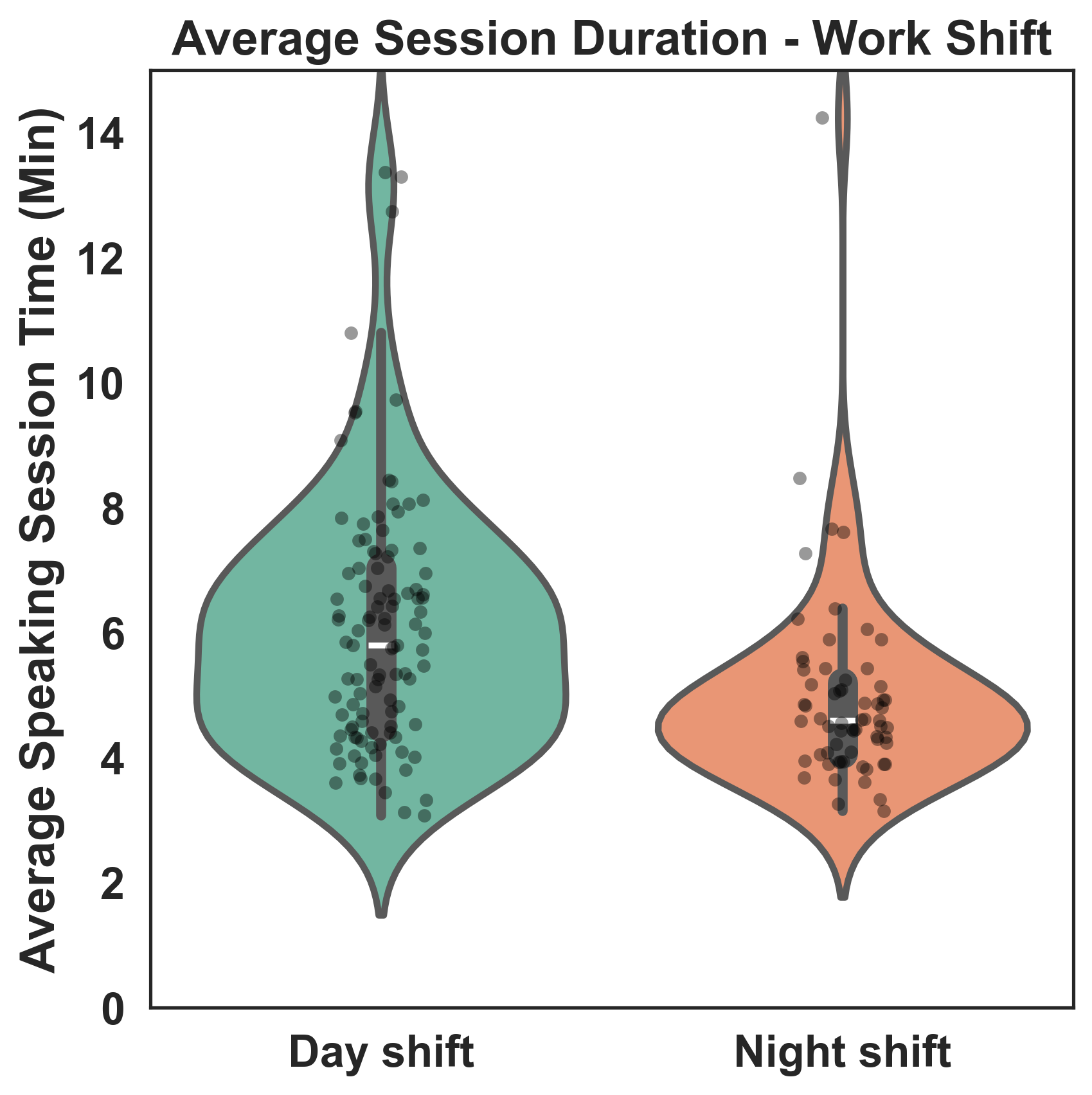}};
        
    \end{tikzpicture}
    \vspace{-6mm}
    \caption{Violin plots comparing speaking session frequency (sessions per hour, left) and average session duration (right) for day-shift and night-shift staff.} 
    \label{fig:session_shift}
    \vspace{-3.5mm}
    
} 
\end{figure}

\subsubsection{Primary Work Shift}

We compared the speaking session frequency per hour and average session duration between day shift and night shift in Figure~\ref{fig:session_shift}. Overall, the violin plots indicated clear shift-level differences in communication patterns. For speaking frequency (sessions per hour), day-shift staff show a higher mean session frequency (Day shift: Mean 3.98, 95\% CI [3.87-4.09], Night shift: Mean 3.63, 95\% CI [3.47-3.79]), with a significant effect of shift (three-way ANOVA, F=12.36, $p<0.001$). We observed a similar pattern for speaking duration, where day-shift speaking sessions are longer on average than night-shift speaking sessions (Day shift: Mean 5.71, 95\% CI [5.43-5.97], Night shift: Mean 4.94, 95\% CI [4.58-5.30]), with a significant effect of shift (three-way ANOVA, F=8.43, $p<0.001$). Together, these results indicate that day-shift staff engaged in more frequent and longer speaking sessions than night-shift staff. This pattern may reflect structural differences between day and night shifts, such as higher patient turnover, increased family interactions, and greater care coordination tasks (e.g., rounds) during the day.

\begin{figure}[ht] {
    \centering
    
    \begin{tikzpicture}
        
        \node[draw=none,fill=none] at (0\linewidth, 4){\includegraphics[width=\linewidth]{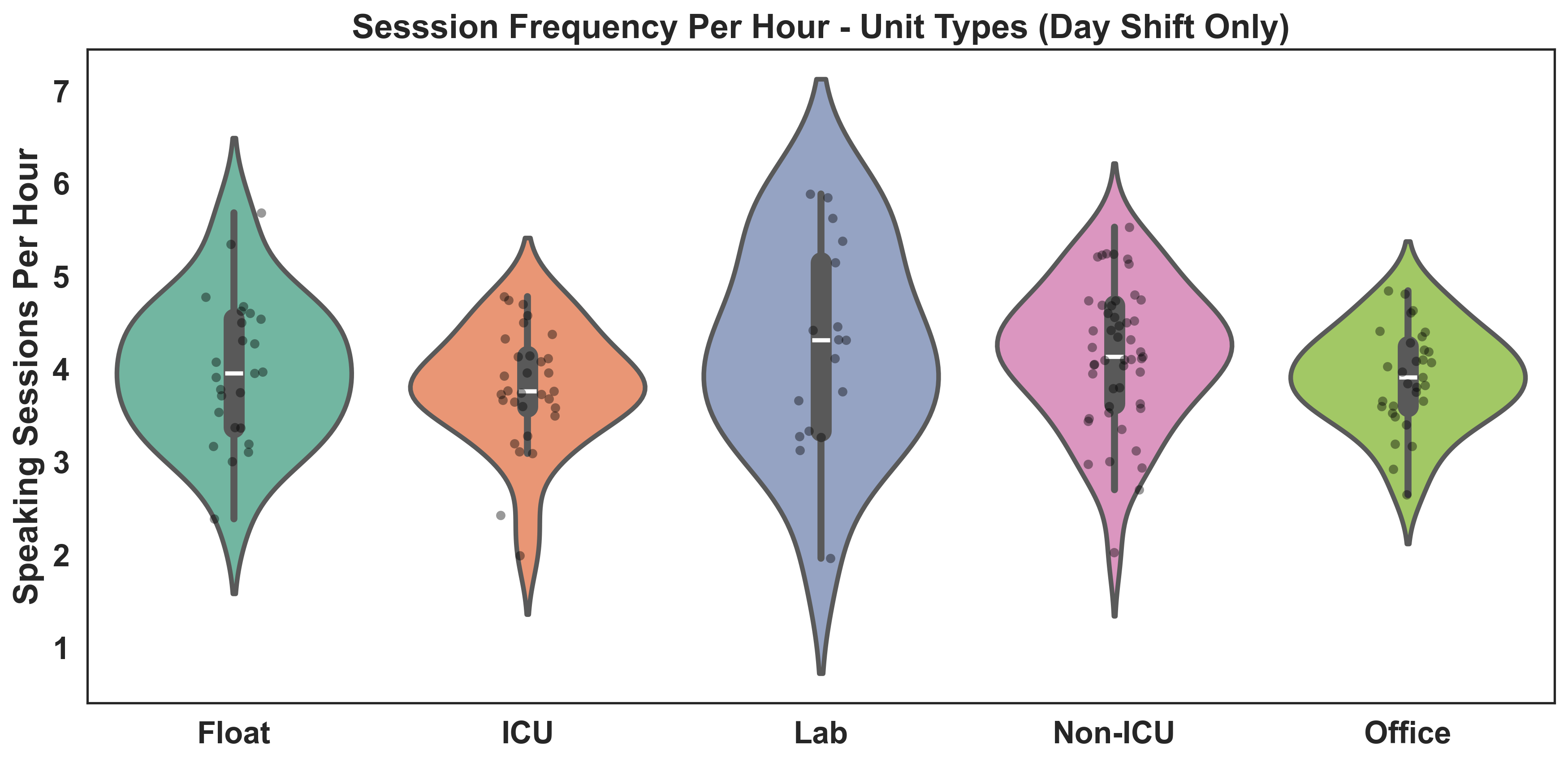}};
        
        \node[draw=none,fill=none] at (0\linewidth, 0){\includegraphics[width=\linewidth]{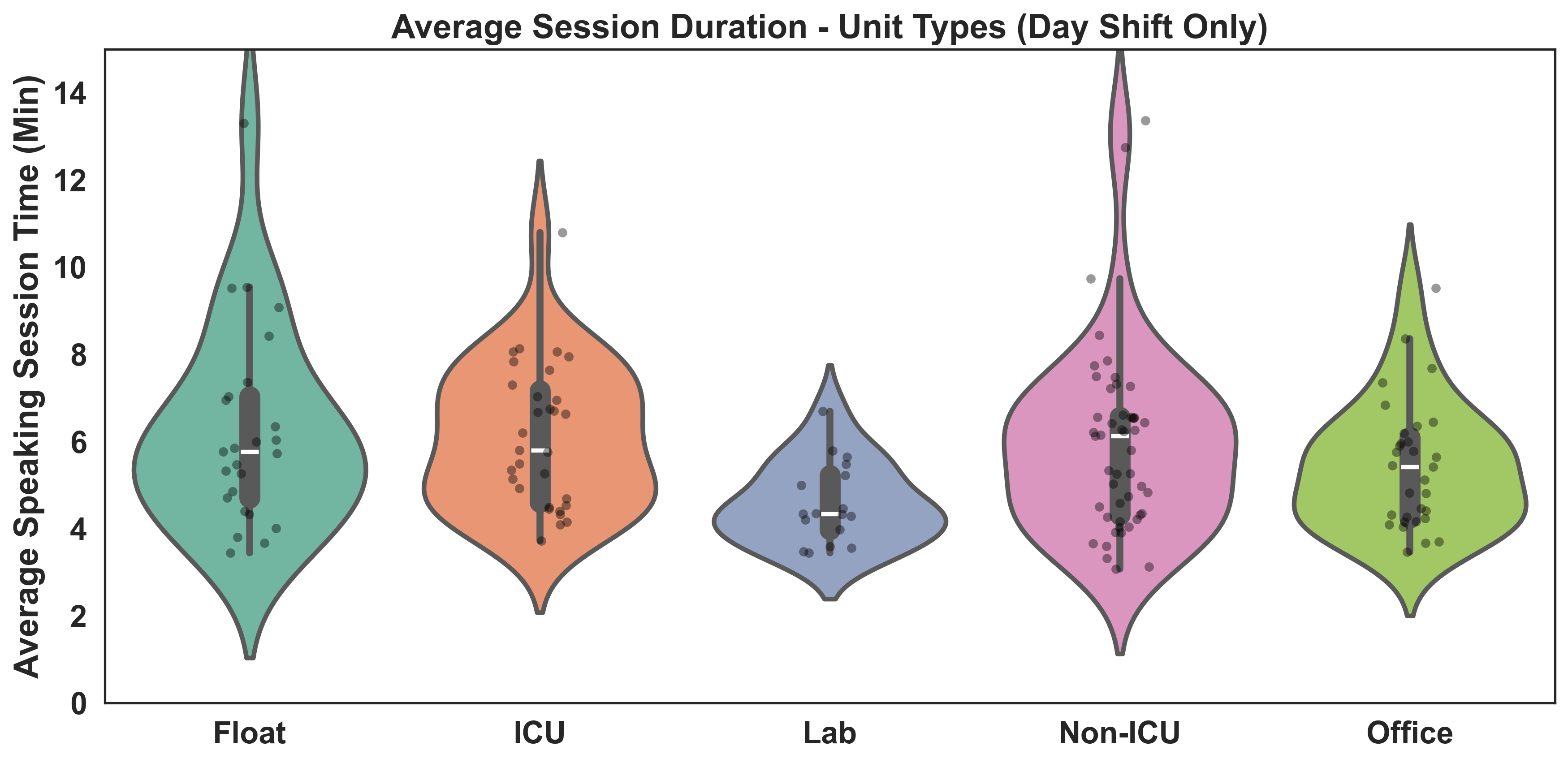}};
        
    \end{tikzpicture}
    \vspace{-6mm}
    \caption{Violin plots comparing speaking session frequency (sessions per hour, top) and average session duration (bottom) across different hospital work units.} 
    \label{fig:session_units}
    \vspace{-3mm}
    
} \end{figure}

\subsubsection{Primary Work Unit}

We further compared communication patterns across work units. Because relatively few subjects worked night shifts in labs and offices, we restricted this comparison to day-shift staff. Figure~\ref{fig:session_units} presents the speaking session frequency per hour and the average speaking session duration across different work units. We observed that staff working in labs engaged in more frequent speaking activity on average, although this difference was not statistically significant. On the other hand, staff working in labs had substantially shorter speaking session durations compared to other work units (Lab: Mean 5.03, 95\% CI [4.48-5.57], ICU: Mean 6.12, 95\% CI [5.51-6.72], Non-ICU: Mean 5.93, 95\% CI [5.32-6.54], Float: Mean 6.25, 95\% CI [5.30-7.20], Office: Mean 5.26, 95\% CI [4.79-5.73]), with a significant effect of work units (three-way ANOVA, F=2.46, $p=0.047$). One explanation is that laboratory work is characterized by frequent task coordination and rapid exchanges (e.g., brief clarifications), which naturally lead to more frequent but shorter speaking sessions. In contrast, nursing units may involve longer, continuous interactions, such as patient communication or care planning, resulting in fewer but longer speaking sessions.

\vspace{-3mm}
\subsection{Speaking Arousal}

We further presented the analysis of the speaking arousal among different work shifts and unit types. We compared the 90th percentile of speaking arousal in the 1st-half and 2nd-half of the work shift. These features describe how active hospital professionals communicate at work. Similar to the previous analysis, we selected only day shift staff when comparing speaking arousal among different working units. We applied a three-way ANOVA in all statistical tests, by adding sex and age as confounding factors.

\subsubsection{Primary Work Shift}

Table~\ref{tab:arousal} highlights systematic differences in speaking arousal between day and night shifts. In the first half of the shift, night-shift staff show higher upper-tail (90th-percentile) arousal than day-shift staff (Day shift: Mean 0.40, 95\% CI [0.38-0.41], Night shift: Mean 0.44, 95\% CI [0.41-0.47]); three-way ANOVA, F=10.34, $p=0.001$). In the second half, this pattern was reversed, where 90th arousal remained relatively stable in day-shift, whereas it decreased in night-shift (Day shift: Mean 0.41, 95\% CI [0.39-0.42], Night shift: Mean 0.35, 95\% CI [0.32-0.38]); three-way ANOVA, F=13.78, $p<0.001$). 
Given this difference, we further plotted changes in 90th Arousal across the shift between day shift and night shift staff working in nursing units in Appendix~\ref{apd:dynamics}. 
As shown, day-shift nurses maintained relatively consistent 90th arousal levels throughout the shift, whereas night-shift nurses showed decreased arousal over time.

Overall, these results suggest that speaking arousal remains relatively stable across the day shift but it declines over the course of the night shift. 
This pattern may reflect circadian misalignment and accumulating fatigue during night work, which can reduce physiological and cognitive activation and, consequently, decrease speaking arousal~\citep{mchill2014impact}.

\begin{table}
    \footnotesize
    \centering

    \vspace{1mm}

    \scalebox{0.86}{
    \begin{tabular}{lccc}

        \toprule
        \multirow{1}{*}{\textbf{}} & 
        \multicolumn{1}{c}{\textbf{90th Arousal}} &
        \multicolumn{1}{c}{\textbf{90th Arousal}} \\ 

        & 
        \multicolumn{1}{c}{\textbf{in 1st-half Shift}} &
        \multicolumn{1}{c}{\textbf{in 2nd-half Shift}} \\

        \midrule

        \multicolumn{1}{l}{\textbf{Primary Work Shift}} & & \\

        \hspace{0.2cm}{Day Shift} &
        0.40 [0.38, 0.41] &
        \textbf{0.41} [0.39, 0.42] \\

        \hspace{0.2cm}{Night Shift} &

        \textbf{0.44} [0.41, 0.47] &
        0.35 [0.32, 0.38] \\

        \hspace{0.2cm}{p-value (Work Shift)} &

        0.001 &
        $<$0.001 \\

        \hspace{0.2cm}{F (Work Shift)} &
        10.34 &
        13.78 \\

        \midrule
        
        \multicolumn{1}{l}{\textbf{Primary Work Unit}} & & \\

        \hspace{0.2cm}{ICU} &
        \textbf{0.46} [0.41, 0.51] &
        \textbf{0.48} [0.42, 0.53] \\

        \hspace{0.2cm}{Non-ICU} &
        0.40 [0.38, 0.41] &
        0.40 [0.39, 0.42] \\

        \hspace{0.2cm}{Float} &
        0.36 [0.31, 0.40] &
        0.40 [0.36, 0.44] \\

        \hspace{0.2cm}{Lab} &
        0.42 [0.38, 0.46] &
        0.40 [0.35, 0.44] \\

        \hspace{0.2cm}{Office} &
        0.37 [0.34, 0.40] &
        0.38 [0.35, 0.41] \\

        \hspace{0.2cm}{p-value (Work Unit)} &

        $<$0.001 &
        0.002 \\

        \hspace{0.2cm}{F (Work Unit)} &
        5.09 &
        4.40 \\

        \bottomrule

    \end{tabular}
    }
    
    \caption{Comparisons of speaking arousal patterns among different work units and shifts. Mean and 95\% CI were reported along with the p-value and F in the three-way ANOVA test.}
    \label{tab:arousal}
    \vspace{-3.5mm}

\end{table}

\vspace{-3mm}
\subsubsection{Primary Work Unit}

\begin{figure}[ht] {
    \centering
    
    \begin{tikzpicture}
        
        \node[draw=none,fill=none] at (0, 3){\includegraphics[width=0.5\linewidth]{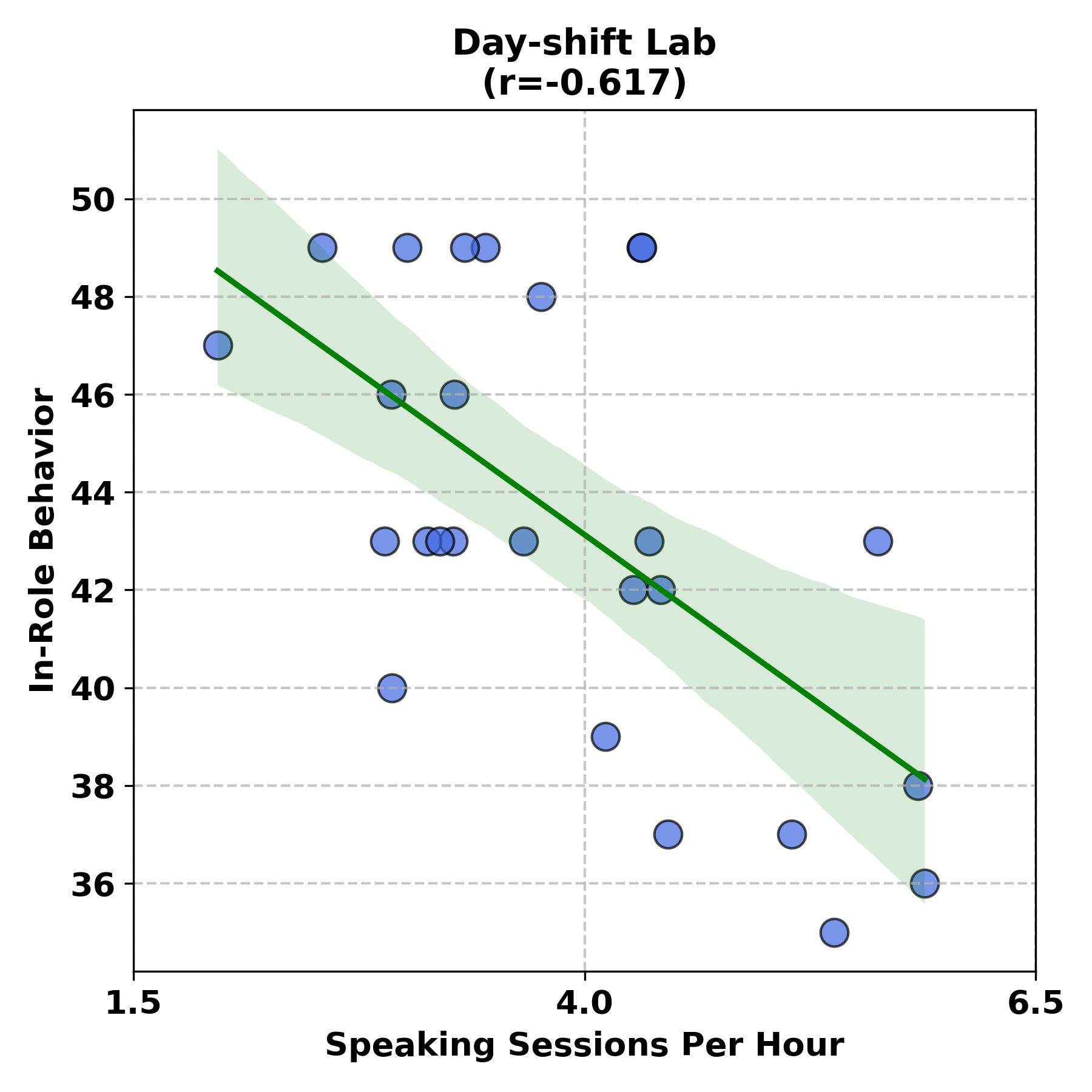}};
        
        \node[draw=none,fill=none] at (0.5\linewidth, 3){\includegraphics[width=0.5\linewidth]{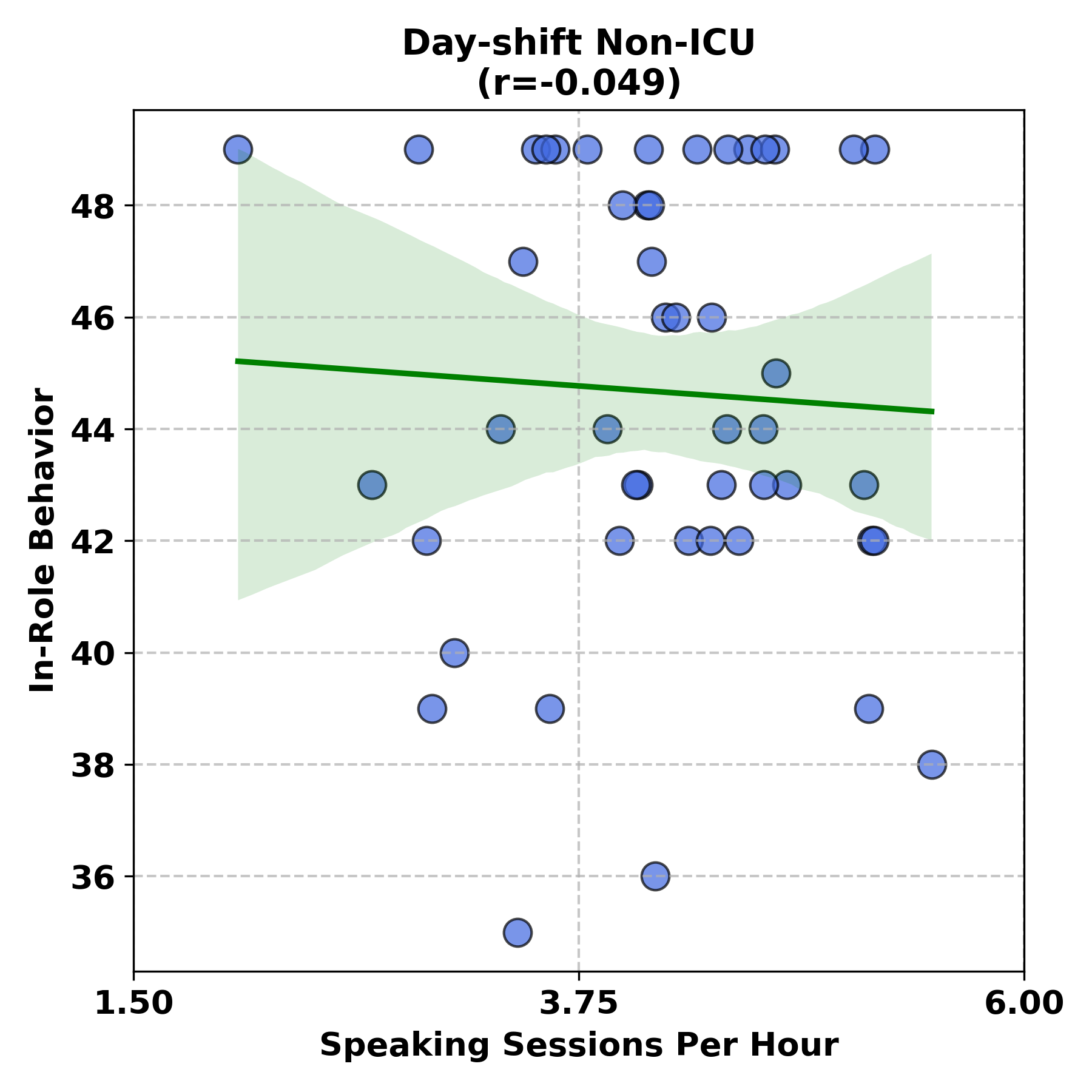}};
        
    \end{tikzpicture}
    \vspace{-7mm}
    \caption{Correlations between speaking frequency and IRB scores among day-shift staff working in labs (left) and Non-ICU units (right).} 
    \label{fig:corr_irb}
    \vspace{-3.5mm}
    
} \end{figure}

Table~\ref{tab:arousal} also reveals systematic differences in speaking arousal across work units. In particular, ICU staff showed higher arousal levels than staff in other units, suggesting more activated communication. This suggests that even though ICU staff have similar speaking frequency and duration compared to other staff, their communication pattern may be characterized more by how they speak (i.e., with greater prosodic activation), potentially reflecting the higher acuity and time-critical nature of ICU workflows. 
In contrast, office staff shows comparatively lower speaking arousal across working hours, aligning with the less interaction-intensive nature of office-based workflows.

\vspace{-3mm}
\section{Communication Patterns and Self-Reported Outcomes}

\subsection{Communication Patterns and Work-Related Outcomes}

Figure~\ref{fig:session_shift} shows a strong negative correlation between speaking session frequency and in-role behavior among day-shift lab staff ($r=-0.617$, $p=0.002$), where more frequent speaking is associated with lower self-reported IRB. One interpretation is that higher communication frequency may reflect greater interruption rates or task switching in lab workflows, potentially reducing the perceived ability to complete core duties. In contrast, among day-shift non-ICU staff, speaking session frequency is uncorrelated with IRB ($r=-0.049$, $p=0.739$), consistent with the intuition that communication may be an expected and integral component of nursing workflows. We further observe that this behavior is consistent for ICU staff and for night-shift staff (see Appendix~\ref{apd:frequency_irb}).

\begin{figure}[ht] {
    \centering
    
    \begin{tikzpicture}
        
        \node[draw=none,fill=none] at (0.5\linewidth, 3){\includegraphics[width=0.5\linewidth]{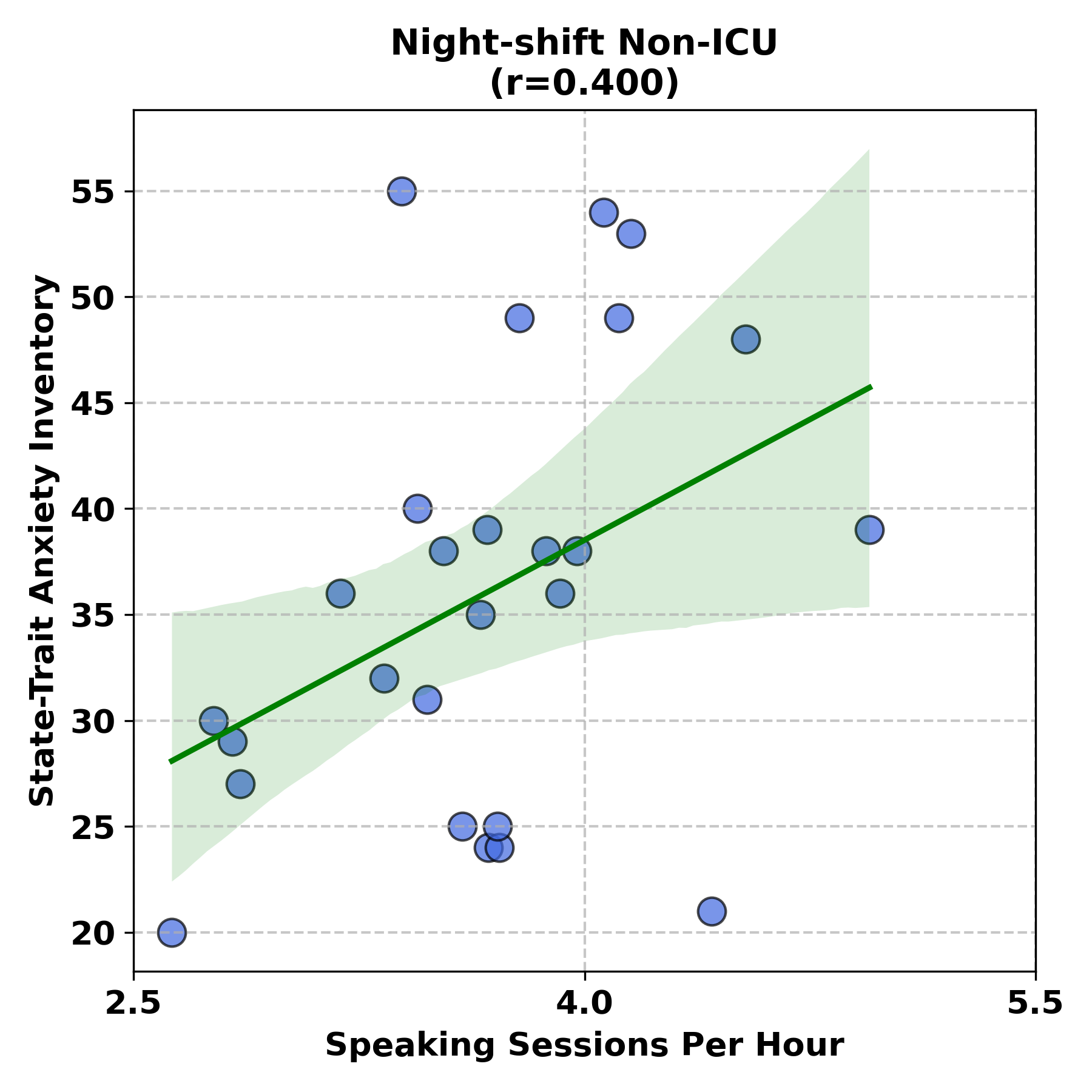}};
        
        \node[draw=none,fill=none] at (0, 3){\includegraphics[width=0.5\linewidth]{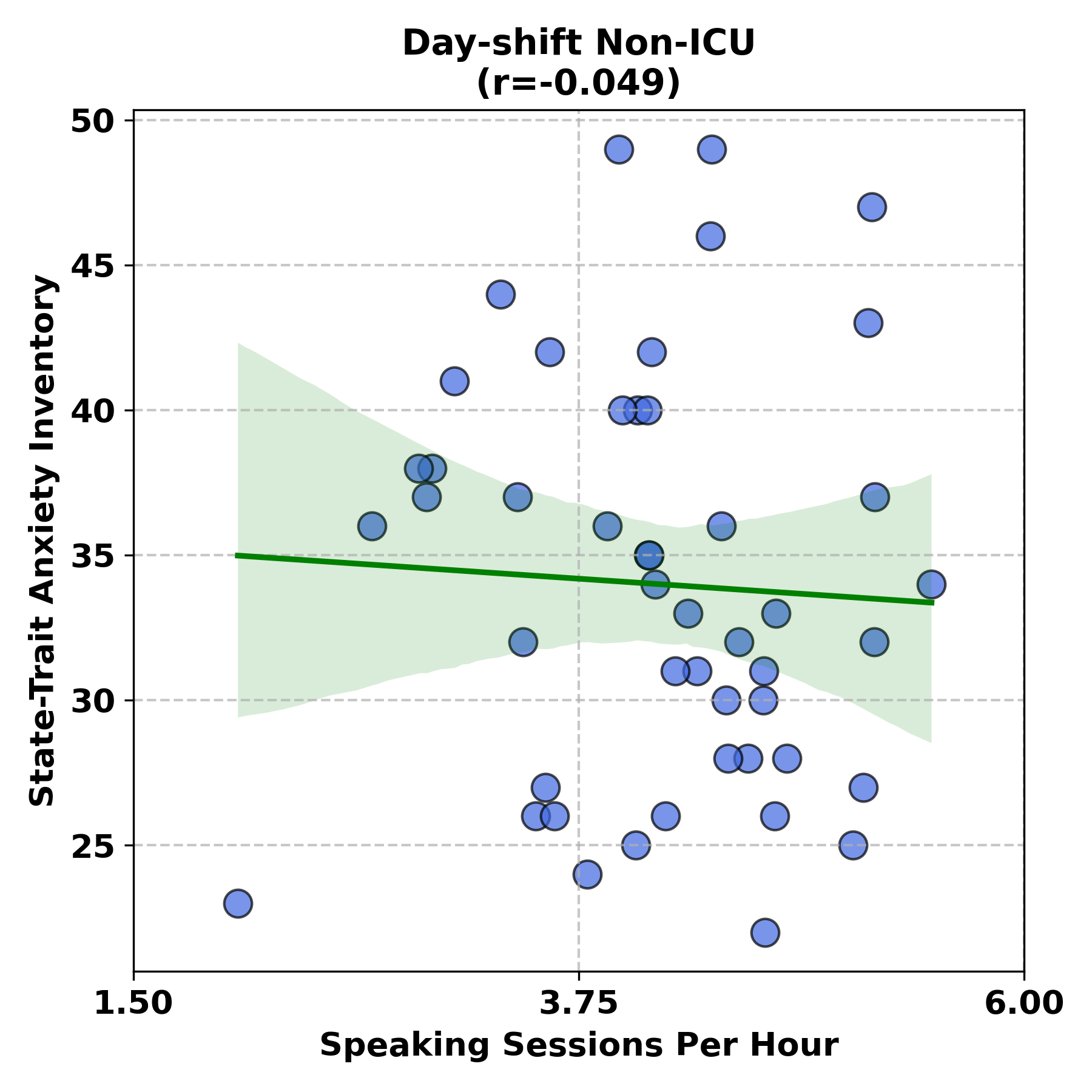}};
        
    \end{tikzpicture}
    \vspace{-7mm}
    
    \caption{Correlations between speaking session frequency and STAI scores among day-shift (left) and night-shift (right) staff working in Non-ICU units.} 
    \label{fig:corr_stai}
    \vspace{-3.5mm}
    
} \end{figure}

\subsection{Communication Patterns and Mental Health Outcomes}

We further explored whether speaking frequency was associated with mental health outcomes measured using STAI. In particular, Fig~\ref{fig:corr_stai} suggests a shift-dependent relationship between speaking session frequency and STAI among non-ICU staff. During the day shift, speaking sessions per hour were uncorrelated with trait anxiety ($r=-0.049$, $p=0.740$), whereas during the night shift, higher speaking frequency was associated with higher trait anxiety ($r=0.400$, $p=0.050$). This may suggest that night-shift nurses used communication to cope with loneliness~\citep{lim2025navigating}, which could contribute to higher STAI scores. However, we did not find any correlations between arousal-based measures and self-reported outcomes, as shown in the Appendix~\ref{apd:arousal_mental}.

\section{Limitations}
\noindent \textbf{Lack of In-domain FG/BG Speaker Diarization Data} Due to privacy and regulatory constraints, researchers generally were unable to collect and annotate large-scale in-domain foreground speech data directly from hospital environments. Although we collected a limited in-domain internal meeting dataset using the same recording device, the FG/BG speaker diarization model was primarily trained on publicly available egocentric and meeting corpora with similar form factors. Consequently, these training data may not fully capture the acoustic complexity and background noise present in real hospital settings.

\vspace{1mm}
\noindent \textbf{Lack of Detailed Work Behavioral Data} Although our outcome measures relied on the validated and widely-used surveys, they failed to provide detailed or task-level measures of job behaviors. Therefore, subject to Institutional Review Board approval, future studies may consider integrating external operational metrics to more precisely validate communication--work behaviors relationships.

\vspace{1mm}
\noindent \textbf{Lack of Semantic Information} \texttt{VoxCare} intentionally excludes the collection of semantic content, but focuses on acoustic features. As a result, the system cannot capture what was said, limiting the interpretation of communicative intent or task context.

\section{Conclusion}

In this work, we introduced \texttt{VoxCare}, a scalable egocentric wearable audio sensing and computing system for capturing natural communication behaviors in real-world occupational settings. By combining on-device acoustic feature extraction, a speech foundation model-guided teacher-student framework for offline FG/BG diarization, and interpretable behavioral feature extraction, \texttt{VoxCare} enabled large-scale analysis of communication patterns in everyday contexts. Particularly, the teacher–student framework provides an effective approach to develop high-performance FG/BG diarization models using only a limited set of acoustic features.

Through a 10-week longitudinal deployment involving 255 hospital professionals, we demonstrated the feasibility of deploying wearable audio sensing in complex, high-stakes clinical settings and revealed systematic differences in communication frequency, duration, and speech arousal across work shifts and unit types. 
Moreover, our analyses uncovered meaningful associations between communication patterns and self-reported work-related and mental health outcomes. For example, in laboratory settings, higher speaking frequency was negatively associated with in-role behavior, suggesting that fragmented communication may interfere with task completion and perceived effectiveness. These results highlight the potential of speech-derived behavioral measures in naturalistic settings as objective indicators of workplace dynamics among hospital professionals, which can inform improvements in healthcare practice.

\section{Acknowledgment}
The research is based upon work supported by the Office of the Director of National Intelligence (ODNI), Intelligence Advanced Research Projects Activity (IARPA), via IARPA Contract No $2017$ - $17042800005$. The views and conclusions contained herein are those of the authors and should not be interpreted as necessarily representing official policies or endorsements, either expressed or implied, of ODNI, IARPA, or the U.S. Government. The U.S. Government is authorized to reproduce and distribute reprints for Governmental purposes, notwithstanding any copyright annotation thereon.

\newpage
\appendix

\section{Diarization Errors}
\label{apd:fg_diarization}

\begin{figure}[h] {
    \centering
    \includegraphics[width=\linewidth]{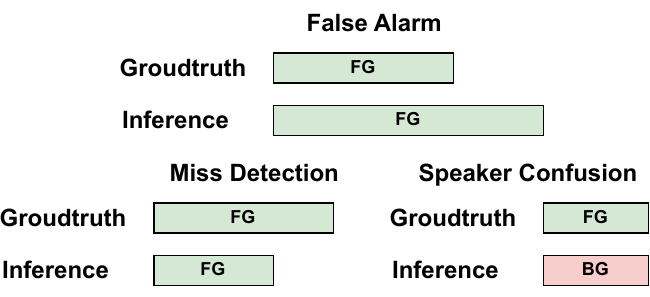} 
    \caption{Illustration of different types of diarization errors used in model evaluation. }
    \label{fig:error}
    \vspace{-3.5mm}
} \end{figure}


We show different types of diarization errors used in model evaluation. Missed Detection refers to the duration of actual speech (FG or BG) in a recording that the system fails to detect, while a false alarm occurs when the system incorrectly identifies non-speech audio, such as background noise or silence, as speech (FG or BG). 
Speaker confusion occurs when the FG or BG speaker is recognized incorrectly as its counterpart.

\section{Survey Descriptions}\label{apd:survey}

\vspace{1mm}
\noindent \textbf{STAI} included 40 self-report items on a 4-point Likert scale regarding state and trait anxiety. The final score was obtained by summing all responses, with higher scores indicating greater anxiety. 

\vspace{1mm}
\noindent \textbf{IRB} describes the extent to which employees report fulfilling the tasks, duties, and responsibilities formally expected as part of their job role. The survey consists of 7 items, each rated on a 7-point Likert scale ranging from 1 (strongly disagree) to 7 (strongly agree). A total score is obtained by summing all item responses, resulting in a score between 7 and 49.

\begin{figure*}[ht] {
    \centering
    
    \begin{tikzpicture}
        
        \node[draw=none,fill=none] at (0\linewidth, 4.3){\includegraphics[width=0.25\linewidth]{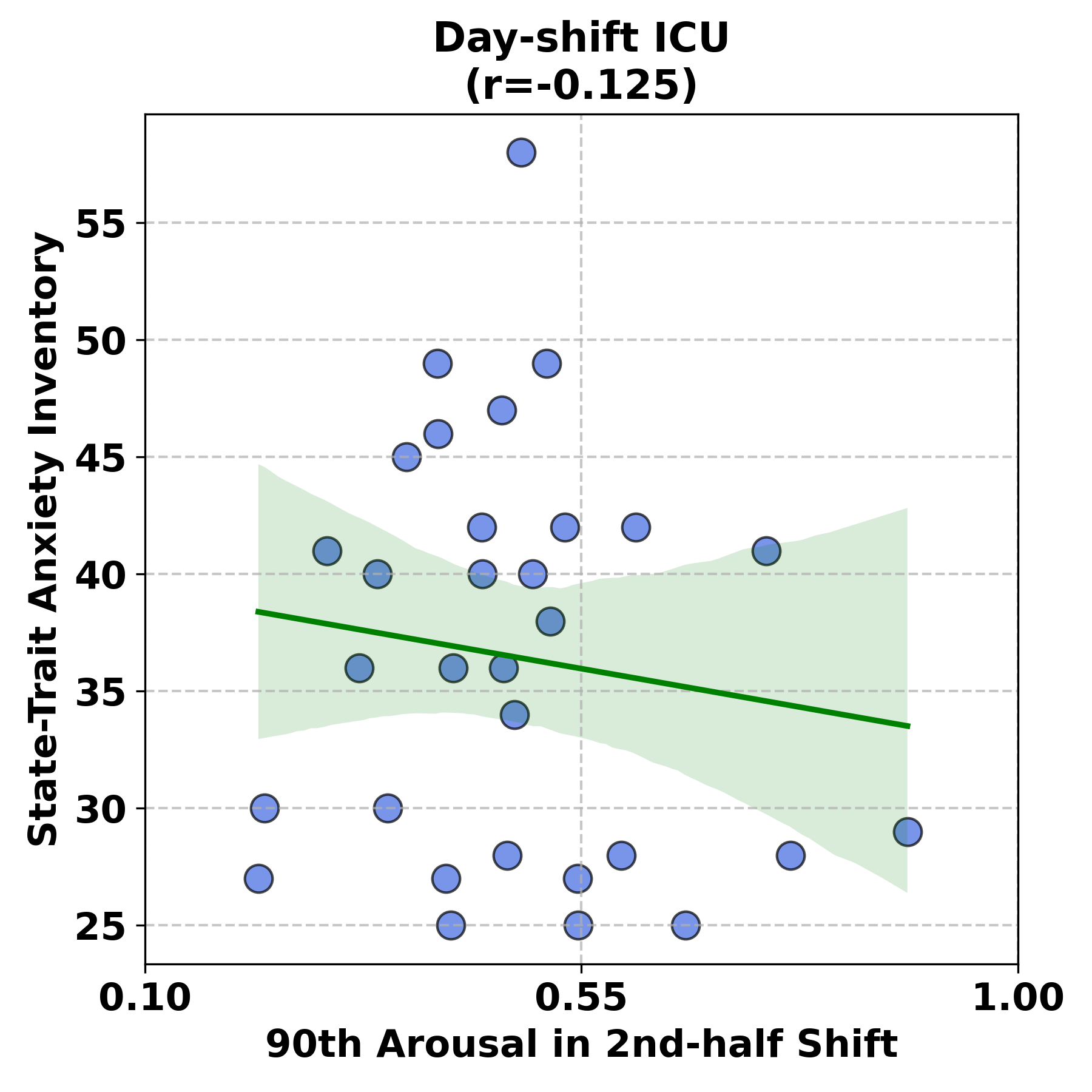}};
        
        \node[draw=none,fill=none] at (0.25\linewidth, 4.3){\includegraphics[width=0.25\linewidth]{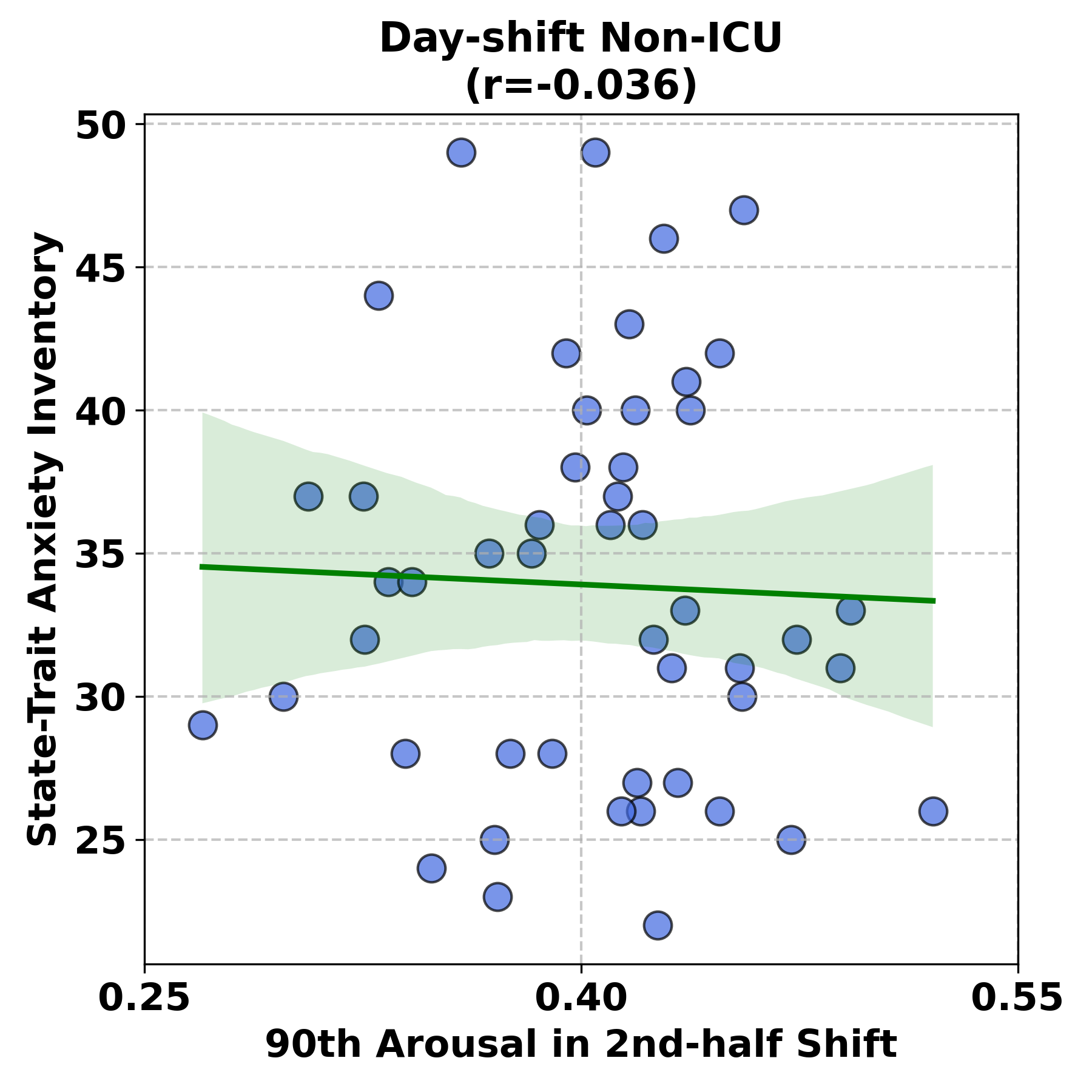}};

        \node[draw=none,fill=none] at (0.5\linewidth, 4.3){\includegraphics[width=0.25\linewidth]{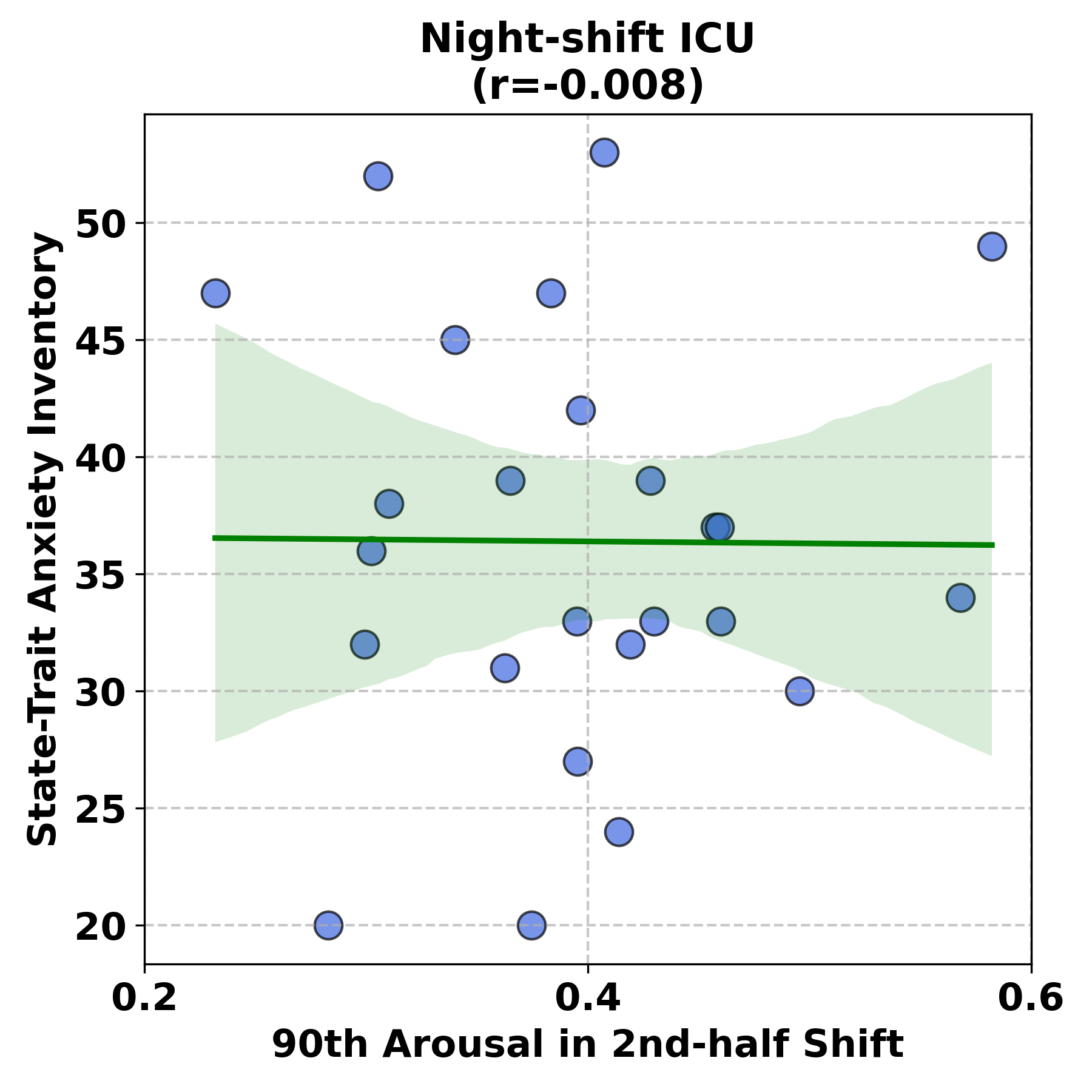}};
        
        \node[draw=none,fill=none] at (0.75\linewidth, 4.3){\includegraphics[width=0.25\linewidth]{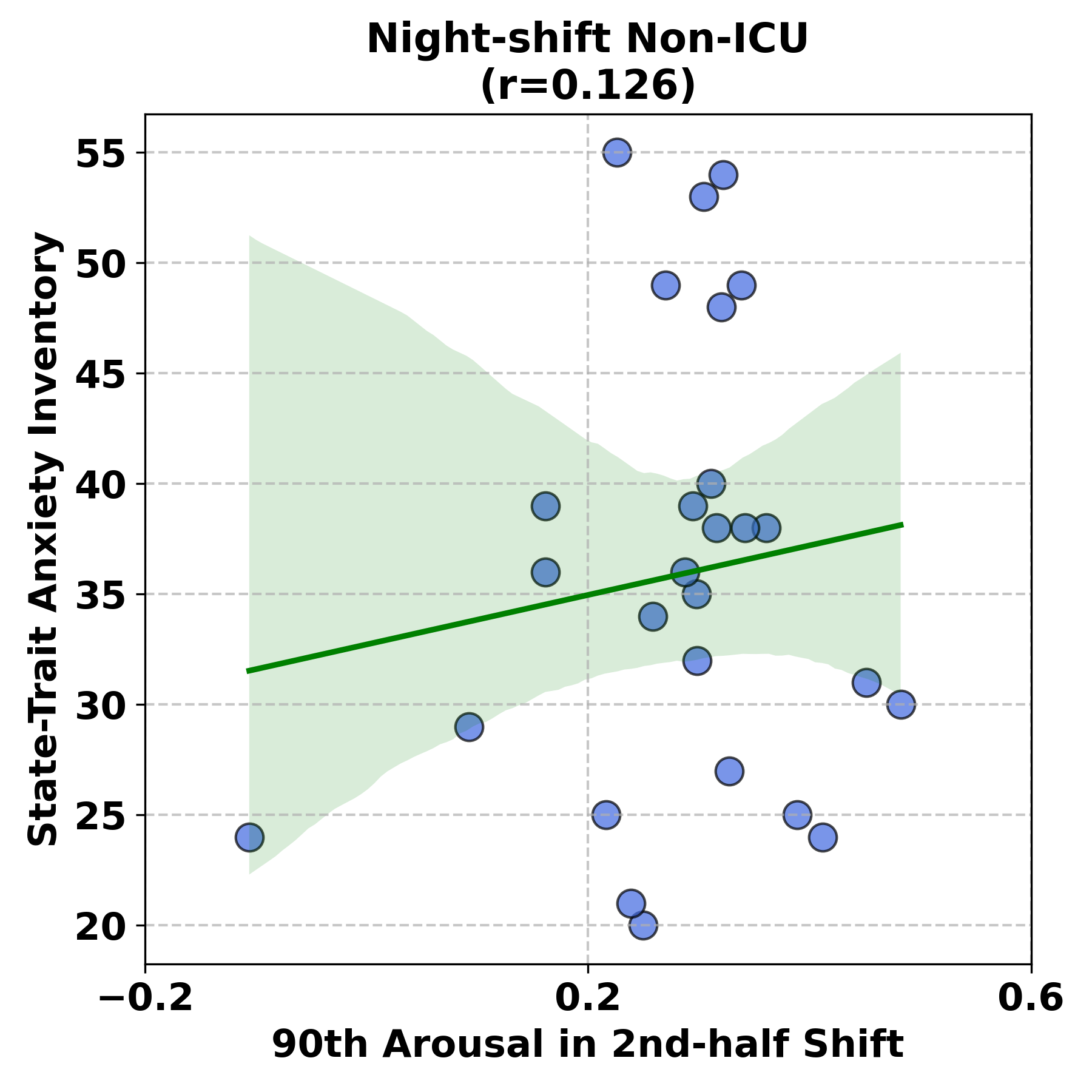}};

        \node[draw=none,fill=none] at (0\linewidth, 0){\includegraphics[width=0.25\linewidth]{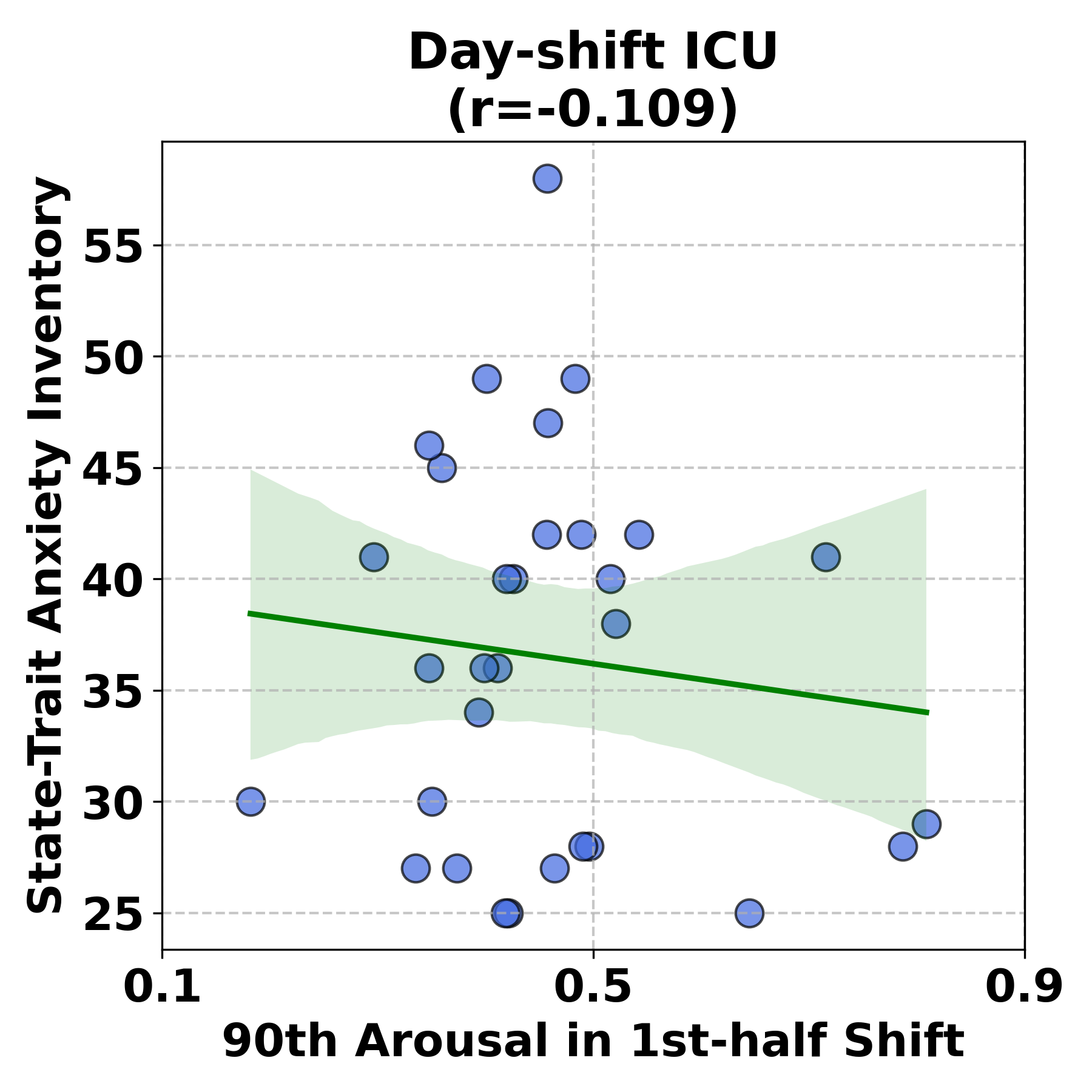}};
        
        \node[draw=none,fill=none] at (0.25\linewidth, 0){\includegraphics[width=0.25\linewidth]{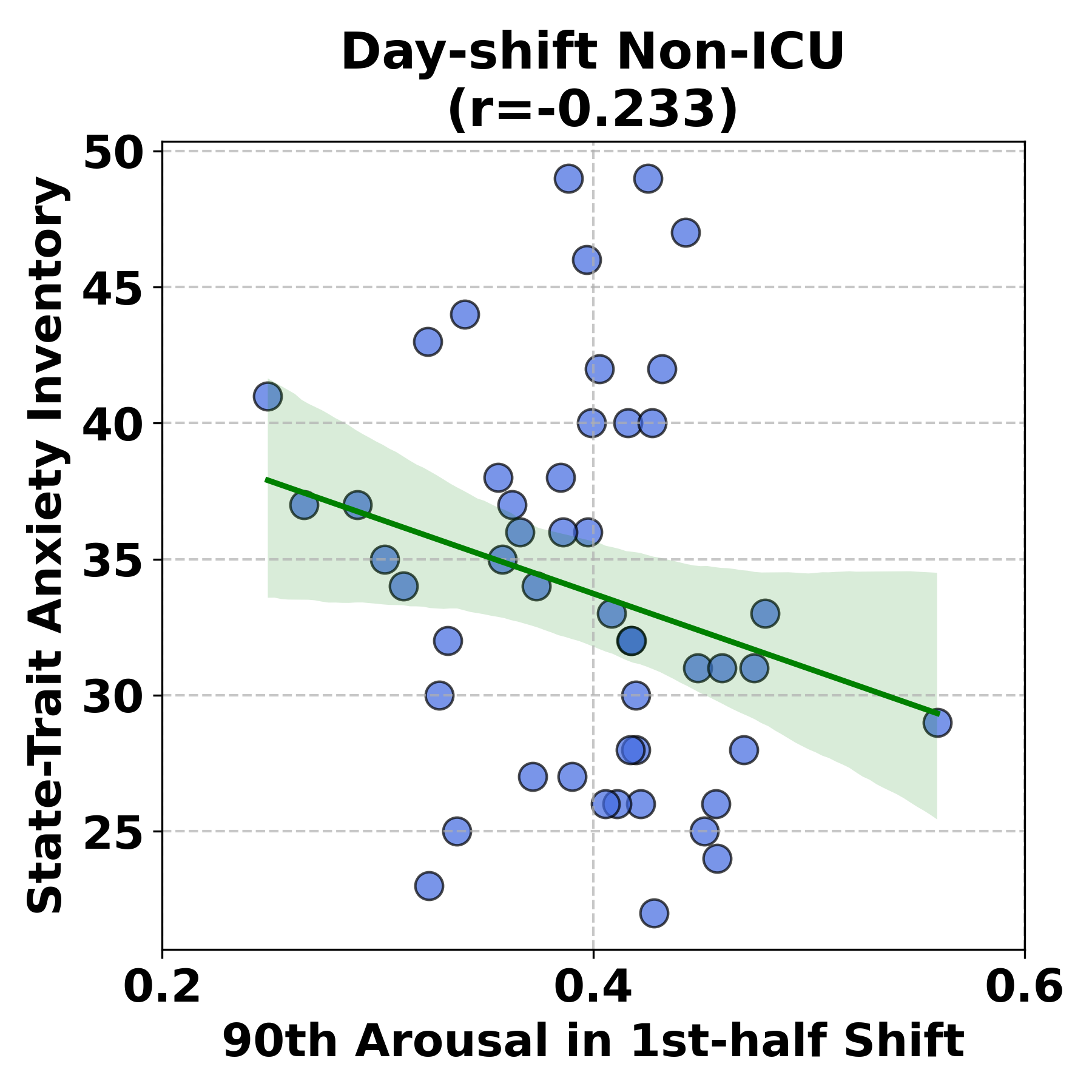}};

        \node[draw=none,fill=none] at (0.5\linewidth, 0){\includegraphics[width=0.25\linewidth]{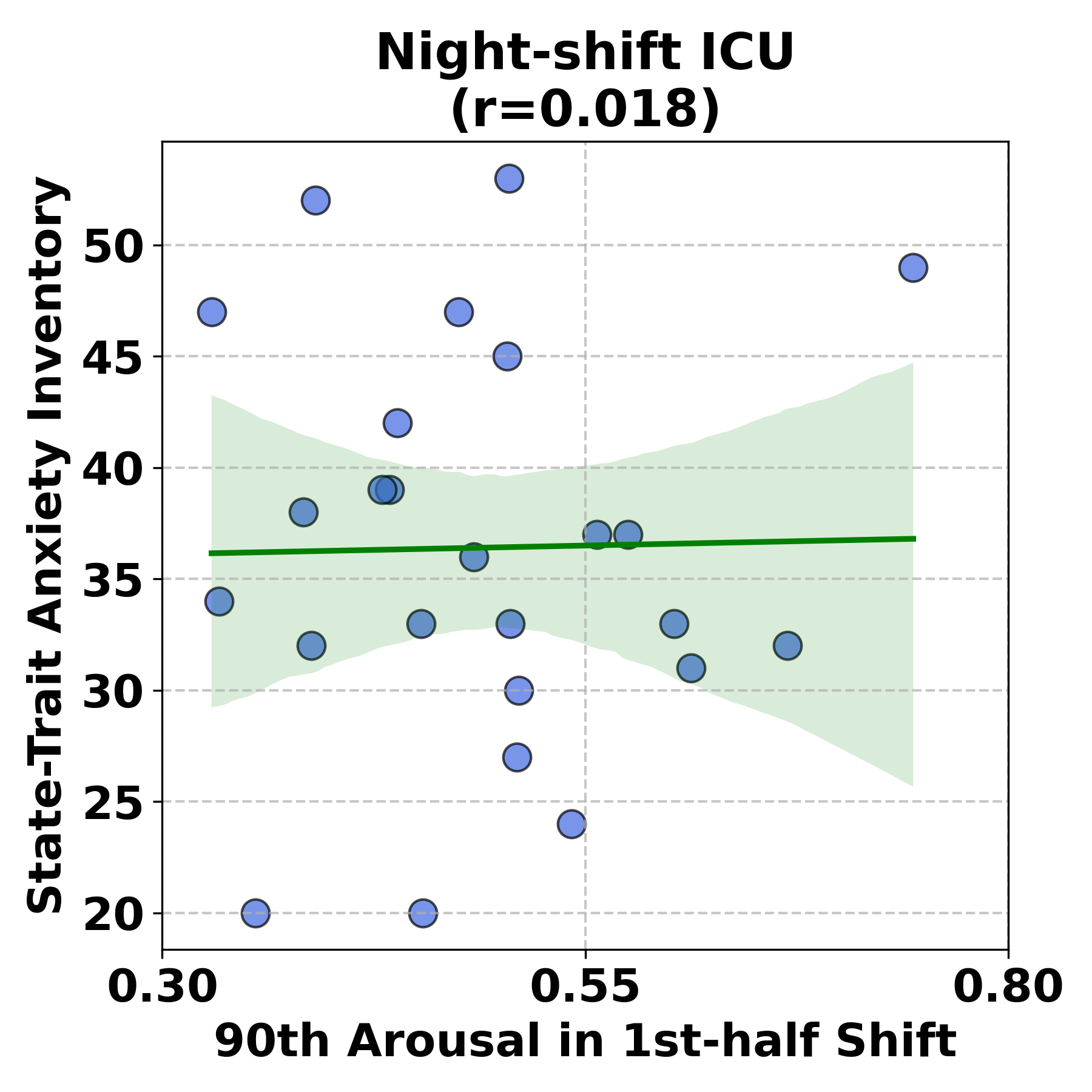}};
        
        \node[draw=none,fill=none] at (0.75\linewidth, 0){\includegraphics[width=0.25\linewidth]{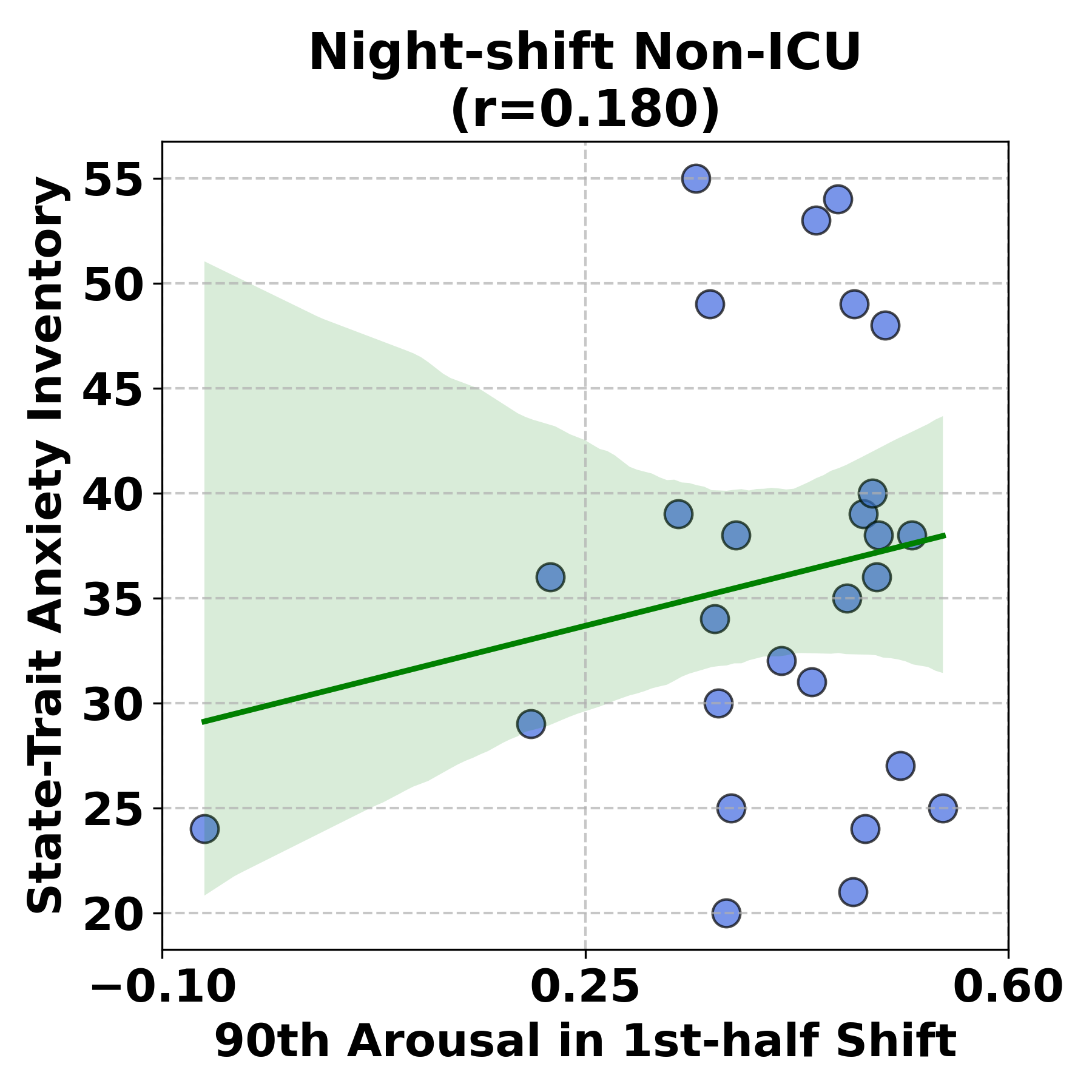}};
        
    \end{tikzpicture}
    \vspace{-7mm}
    
    \caption{Correlations between 90th percentile speaking arousal and STAI scores among day-shift and night-shift staff working in Non-ICU and ICU units.} 
    \label{fig:corr_stai_app}
    \vspace{-3.5mm}
    
} \end{figure*}

\section{Dynamics of 90th Arousal between Shift}\label{apd:dynamics}

\begin{figure}[ht] {
    \centering
    {\includegraphics[width=\linewidth]{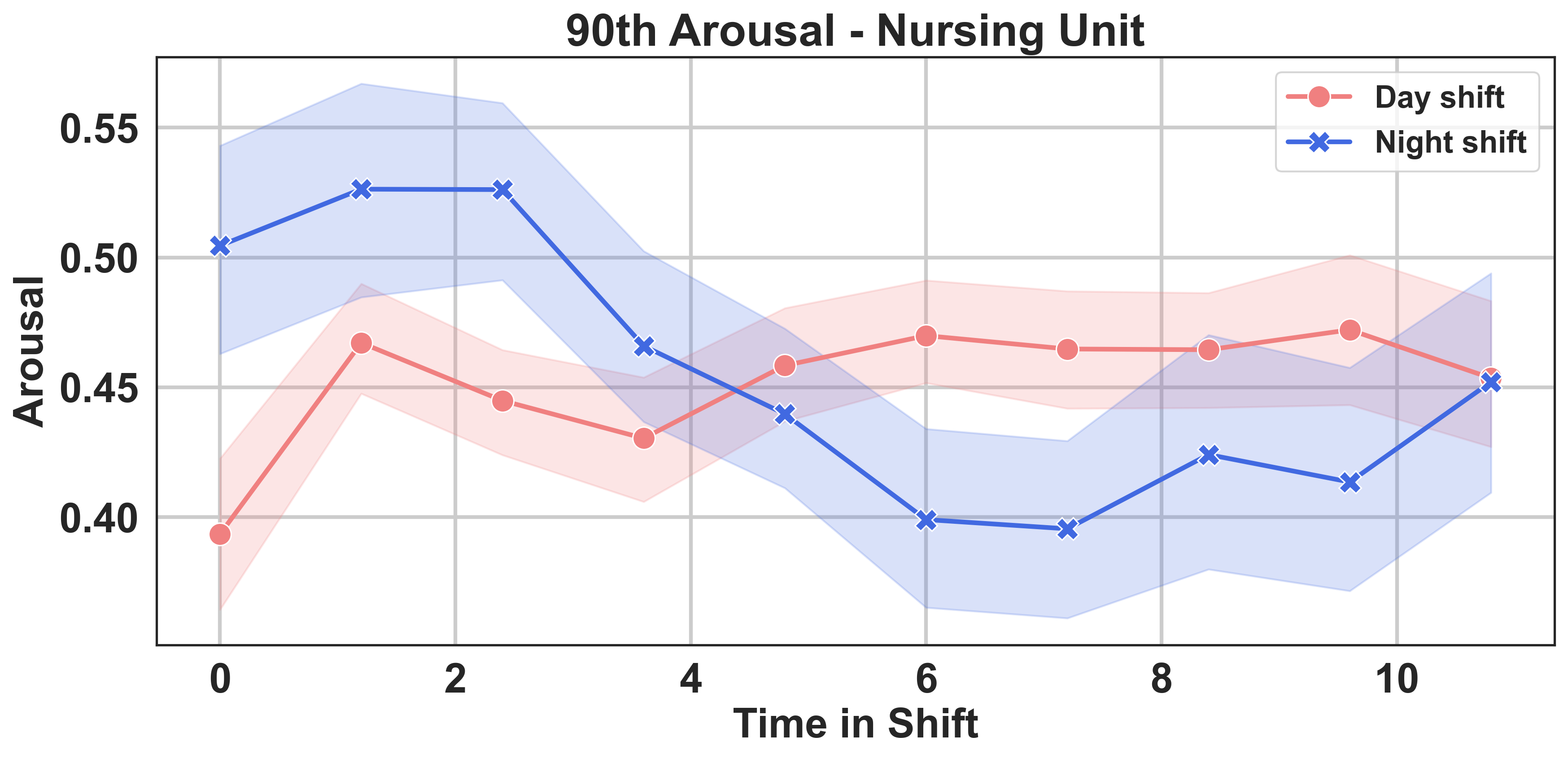}}
    \vspace{-6mm}
    \caption{Changes in 90th Arousal across the shift between day shift and night shift staff working in nursing units.} 
    \label{fig:arousal_shift}
    \vspace{-3mm}
    
} \end{figure}

Here, we present changes in 90th arousal across the shift between day shift and night shift staff working in nursing units in Figure~\ref{fig:arousal_shift}.

\section{Speaking Frequency and In-role Behavior}\label{apd:frequency_irb}

We show the correlations between speaking frequency and IRB scores among night-shift staff working in Non-ICU and ICU units. The plots indicate no significant correlations between speaking frequency features and IRB scores.

\begin{figure}[ht] {
    \centering
    
    \begin{tikzpicture}
        
        \node[draw=none,fill=none] at (0.5\linewidth, 3){\includegraphics[width=0.5\linewidth]{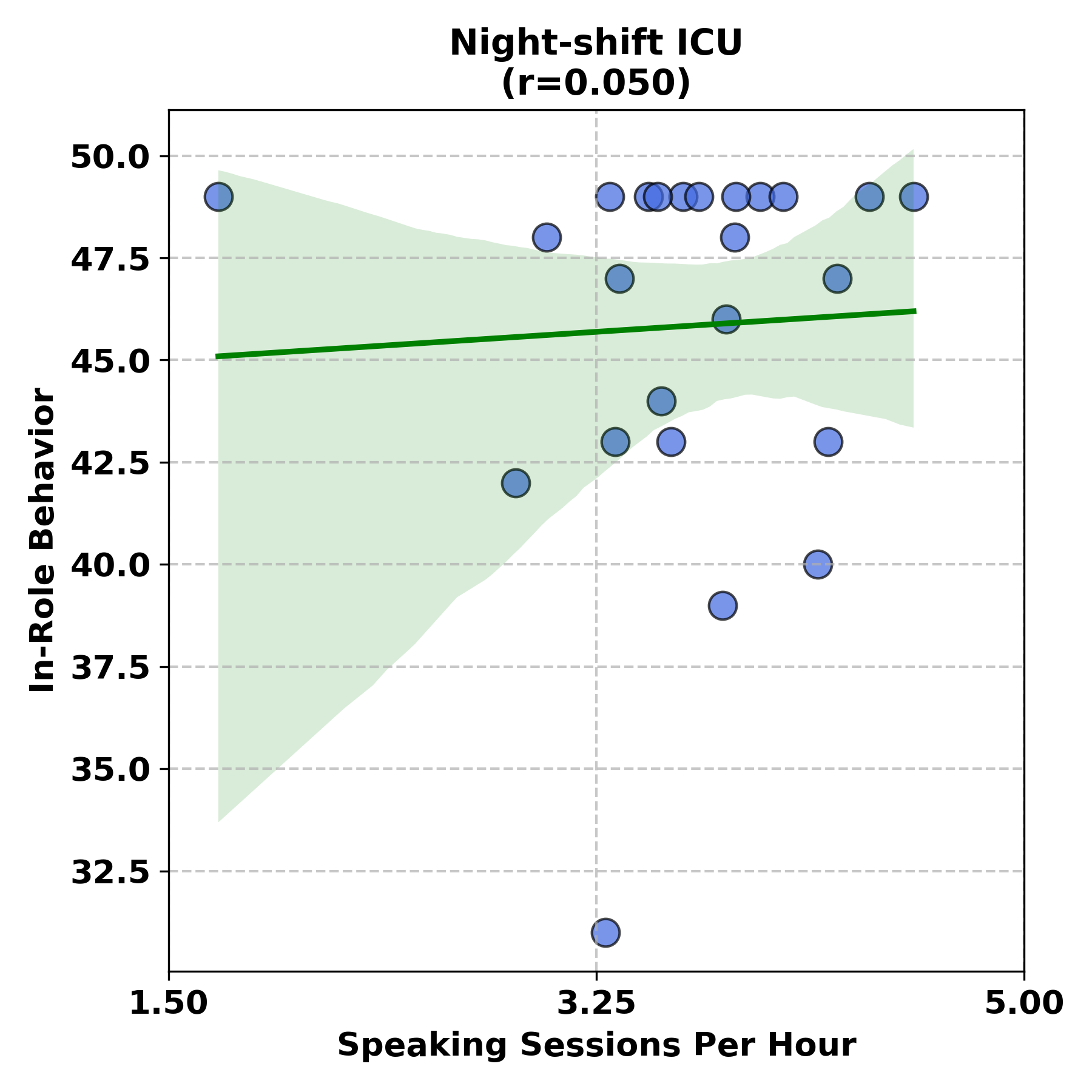}};
        
        \node[draw=none,fill=none] at (0, 3){\includegraphics[width=0.5\linewidth]{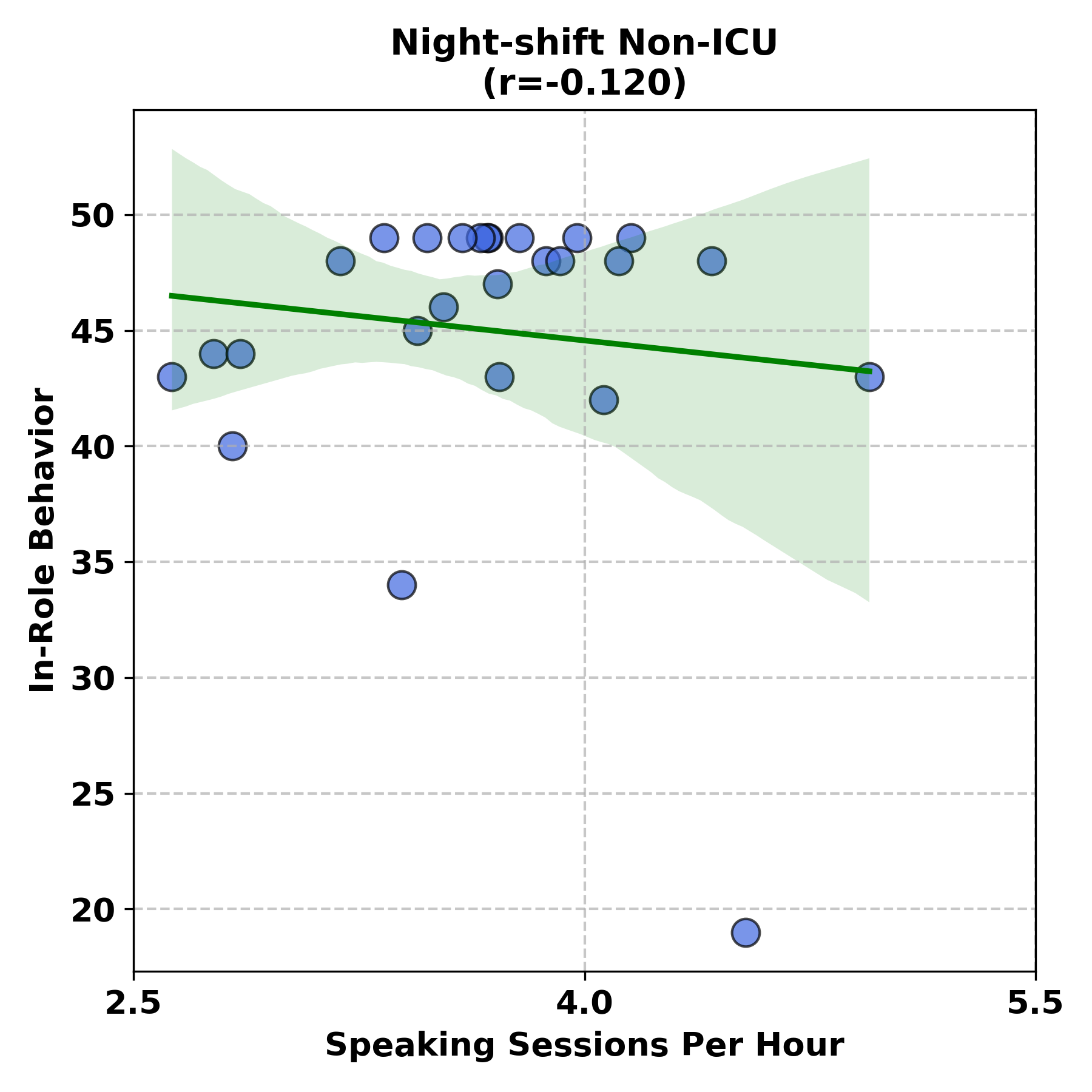}};
        
    \end{tikzpicture}
    \vspace{-7mm}
    
    \caption{
    Correlations between speaking frequency and IRB scores among night-shift staff working in ICU (left) and Non-ICU units (right).} 
    \label{fig:corr_irb_app}
    \vspace{-3.5mm}
    
} \end{figure}

\section{Speaking Arousal and Mental Health Outcomes}\label{apd:arousal_mental}

We show the correlations between 90th percentile speaking arousal and STAI scores among day-shift and night-shift staff working in Non-ICU and ICU units in Figure~\ref{fig:corr_stai_app}. The plots indicate no significant correlations between speaking arousal features and STAI scores.

\newpage
\newpage
\bibliography{refs}

\end{document}